\documentclass[%
 aip,
 amsmath,amssymb,
reprint,%
]{revtex4-1}

\usepackage{graphicx}% Include figure files
\usepackage{dcolumn}% Align table columns on decimal point
\usepackage{bm}% bold math
\usepackage{xr}

\makeatletter
\newcommand*{\addFileDependency}[1]{% argument=file name and extension
  \typeout{(#1)}
  \@addtofilelist{#1}
  \IfFileExists{#1}{}{\typeout{No file #1.}}
}
\makeatother

\newcommand*{\myexternaldocument}[1]{%
    \externaldocument{#1}%
    \addFileDependency{#1.tex}%
    \addFileDependency{#1.aux}%
}

\myexternaldocument{SI}

\usepackage[utf8]{inputenc}
\usepackage[T1]{fontenc}
\usepackage{mathptmx}
\usepackage{color}
\usepackage[table, x11names]{xcolor}
\usepackage{appendix}
\usepackage{braket}

\usepackage{comment}
\usepackage{longtable}
\usepackage{etoolbox}
\usepackage{hyperref}
\hypersetup{
    colorlinks=true,
    linkcolor=blue,
}

\usepackage{pgffor}

\begin{document}

\preprint{AIP/123-QED}

\title[]{
Studying excited-state-specific perturbation theory
on the Thiel set
}
% Force line breaks with \\

\author{Rachel Clune}
\thanks{These authors contributed equally to this work.}
\affiliation{Department of Chemistry, University of California, Berkeley, California 94720, USA}
\author{Jacqueline A. R. Shea}
\thanks{These authors contributed equally to this work.}
\affiliation{VeriSIM Life, San Francisco, California 94104, USA}
\author{Tarini S. Hardikar} %TODO: need to modify author list in the submission as well!!
\affiliation{Department of Chemistry, University of California, Berkeley, California 94720, USA}
\author{Harrison Tuckman}
\affiliation{Department of Chemistry, University of California, Berkeley, California 94720, USA}
\author{Eric Neuscamman}
\email{eneuscamman@berkeley.edu}
\affiliation{Department of Chemistry, University of California, Berkeley, California 94720, USA}
\affiliation{Chemical Sciences Division, Lawrence Berkeley National Laboratory, Berkeley, CA, 94720, USA}

\date{\today}% It is always \today, today,

\begin{abstract}

We explore the performance of a recently-introduced
$N^5$-scaling excited-state-specific second order
perturbation theory (ESMP2) on the
singlet excitations of the
Thiel benchmarking set.
We find that, without regularization,
ESMP2 is quite sensitive to $\pi$ system size,
performing well in molecules with small $\pi$ systems
but poorly in those with larger $\pi$ systems.
With regularization, ESMP2 is far less sensitive to
$\pi$ system size and shows a higher overall accuracy
on the Thiel set than CC2, EOM-CCSD,
CC3, and a wide variety of time-dependent
density functional approaches.
Unsurprisingly, even regularized ESMP2 is
less accurate than multi-reference perturbation
theory on this test set, which can in part be explained
by the set's inclusion of some doubly excited states
but none of the strong charge transfer states that often pose
challenges for state-averaging.
Beyond energetics, we find that the ESMP2 doubles norm
offers a relatively low-cost way
to test for doubly excited character without
the need to define an active space.

%The Thiel benchmarking set for vertical excitations was used to benchmark second-order excited state M\o ller-Plesset Perturbation theory (ESMP2) and a regularized version of ESMP2, $\epsilon$-ESMP2. The results are compared to CC2, CC3, EOM-CCSD, SA-CASPT2, and MS-CASPT2 results on the same set of excited states and broad comparisons are made to TD-DFT results. We found that while ESMP2 is fairly inaccurate for this set, except for molecules with very little conjugation, adding a level shift of 0.5 Ha dramatically reduced the average error of the method making $\epsilon$-ESMP2 similar in accuracy to CC2 and CC3. A diagnostic was also proposed based on the norm of the amplitudes of the first order ESMP wave function as it should be a measure of the amount of singly excited character in a given excited state. We show that is measure does correspond well to the magnitude of the ESMP2 error and correlates with other measures of excited state character. 

\end{abstract}

\maketitle

\section{Introduction}

%\textcolor{red}{Paragraph 1:  Cost challenges with available single-ref methods; TDDFT; end paragraph on LR/EOM CC methods; note success of CC2 and its N5 cost.}
%
%\textcolor{red}{Paragraph 2:  ESMF, ESMP2}
%
%\textcolor{red}{Paragraph 3:  Thiel set very broad summary and references to the many studies that have employed it.}
%
%\textcolor{red}{Paragraph 4:  Thiel set categories.}
%
%\textcolor{red}{Paragraph 5:  Summary of our findings.}

Quantum chemistry approaches to modeling singly excited states have been
highly successful, but it remains true that the methods that are most
reliably accurate are also highly computationally intensive.
As in ground state theory, coupled cluster (CC) methods that go beyond
doubles but stop short of a full treatment of triples are often
used as reliable benchmarks.\cite{ariyarathna_benchmark,beizaei_benchmark,dral_benchmark,jorgensen_benchmark,pedersen_benchmark,piecuch_benchmark,sauer_soppa_benchmark,sharma_benchmark,tajti_benchmark,tuna_benchmark}
However, with a cost that scales as $N^7$ with the system size $N$,
these methods are quite limited in the size of molecule that they can treat.
Density functional theory (DFT), and in particular time-dependent DFT (TD-DFT),
is much more affordable, with costs ranging from $N^3$ to $N^5$ depending on the
functional, with $N^4$ being typical for many hybrid functionals.
By choosing a functional that is known to work well for the chemistry in question, 
TD-DFT can offer impressive accuracy, especially for its computational price,
but it would be an overstatement to claim that it is as reliable as CC
methods that include some triples effects.
Lower cost CC options for excited states, especially in the linear response (LR)
and equation of motion (EOM) formalisms, are also widely used, but without
triples effects, these methods are more varied in their reliability.
Examples include EOM-CC with singles and doubles (EOM-CCSD),
\cite{krylov:eomccsd}
which has an $N^6$
cost but tends to overestimate excitation energies, and CC2,
\cite{christiansen1995cc2,sneskov2012-WIRES}
which has an $N^5$ cost and typically displays lower average errors than
EOM-CCSD.
These methods are both widely used
and have been quite successful, but nonetheless there is
room for improvement, as they can produce surprisingly large errors
in some cases that are not obviously ill-suited to their assumptions,
as in the 2$^1$A$'$ state of formamide.
Adding partial triples contributions -- as in
CC3, \cite{christiansen1995cc3,koch1997cc3,sneskov2012-WIRES}
EOM-CCSD(T), \cite{eom_ccsd_paren_t,benchmarking_ee_eom}
$\delta$-CR-EOM(2,3)D, \cite{piecuch:delta_eom_ccsd}
and many related methods --
can certainly improve matters, but brings us back to $N^7$ scaling.
In this study, we will use a large test set to investigate to what
degree it may be helpful to move away from the linear response
paradigm and instead build traditional correlation methods upon
a mean field reference, starting, for now, with
second order perturbation theory.

Like ground state second order M{\o}ller-Plesset
perturbation theory (MP2), \cite{Szabo-Ostland}
the recently introduced excited-state-specific 
M{\o}ller-Plesset theory (ESMP2)
\cite{Shea-ESMF-2018,Shea-GVP-2020,Clune-ESMP2-2020}
seeks to provide a second order Rayleigh-Schr{\"o}dinger
correction atop a mean field starting point.
In the ground state, MP2 perturbs around Hartree-Fock theory,
while in ESMP2 the starting point is provided by excited
state mean field (ESMF) theory,
\cite{Shea-ESMF-2018,Shea-GVP-2020}
which refines the configuration interaction singles (CIS)
picture \cite{HeadGordon:2005:tddft_cis}
through excited-state-specific orbital relaxations
to create a method that shares much in common with ground
state mean field theory. \cite{hardikar2020self}
The early studies of ESMP2 have shown promising accuracy,
which has become more relevant thanks to a refinement
of the theory \cite{Clune-ESMP2-2020}
that brings its cost scaling down to $N^5$.
This is asymptotically comparable to MP2, although it should
be noted that ESMP2's cost is an iterative $N^5$ due
to its zeroth order Hamiltonian not being diagonal.
With a relatively low scaling and early promise in
initial tests, we now seek to deepen our understanding
of the strengths and weaknesses of ESMP2 by exploring
its performance on a widely used excited state
benchmark.

The Thiel set \cite{Thiel-BM-MAIN} offers theoretical best estimates (TBEs)
for over one hundred singlet excited states
(and also many triplet states)
spread over 28 molecules,
which include nucleobases, carbonyls, aromatic rings, heterocyclic rings,
small polyenes, and other small unsaturated hydrocarbons.
In the original work, both complete active space second-order
perturbation theory (CASPT2) \cite{roos1982caspt2,andersson1992caspt2}
as well as the LR or EOM coupled cluster methods CC2, EOM-CCSD, and CC3
were compared across these molecules.
Since then, a large number of other research groups have
used the Thiel set to make further comparisons between methods.
\cite{laurent2013tddft,bannwarth_benchmark,battaglia_benchmark,feldt_benchmark,haldar_benchmark,hodecker_benchmark,holzer_benchmark,hubert_behcmark,kempfer_benchmark,kollmar_benchmark,levi_benchmark,mckeon_benchmark,neville_benchmark,pernal_benchmark,sauer_benchmark,song_benchmark,Thiel_review}
Both the quality of the initial test set and its broad
subsequent use make the Thiel set especially attractive for helping to
put ESMP2 in context and for understanding its strengths and weaknesses.
We note that, for consistency with this significant body of previous work,
we have employed the original test set's TBEs in our analysis below,
although we acknowledge that in some cases, such as the nucleobases,
\cite{CC_Thiel_benchmark,kannar2014-nucleobases}
more recent studies may offer even more reliable best estimates.

Thiel and coworkers organized their test set into four groups of molecules.
In one group they placed aldehydes, ketones, and amides, in which, with the exception of
benzoquinone, ESMP2 shows a highly competitive performance even without regularization.
Another group contains unsaturated aliphatic hydrocarbons, some of whose excited states
have large amounts of doubly excited character and so do not satisfy the
assumptions of ESMP2's singly-excited zeroth order reference state.
Although ESMP2 cannot treat doubly excited character accurately, it does prove to be
a relatively cost-effective way to offer warning that such character is present.
Thiel's third group consists of aromatic rings and heterocycles, in which ESMP2's
sensitivity to $\pi$ system size and the practical efficacy of
regularization become especially apparent.
In our discussion below, we reorganize these two groups into three
-- conjugated polyenes, heterocycles, and other rings --
as the polyenes are particularly illuminating for ESMP2.
The fourth and final Thiel group contains the nucleobases
cytosine, thymine, uracil, and adenine.
As in other cases, ESMP2 struggles with their $\pi$ system sizes but improves
substantially with regularization.

\section{Theory}

\subsection{Zeroth Order Reference}
\label{sec:zeroth}

The zeroth order reference state for ESMP2 is the ESMF wave function, which
in its simplest form is an orbital-relaxed linear combination of
single excitations that can be written as follows.
\begin{align}
    %\ket{\Phi_0} = e^{\hat{X}}\left(c_0\ket{\phi_0} + \sum_{ia} c_{ia}\ket{_i^a} + \sum_{\bar{i}\bar{a}}c_{\bar{i}\bar{a}}\ket{_{\bar{i}}^{\bar{a}}}\right)
    \ket{\Phi_0} = e^{\hat{X}}\left( \sum_{ia} c_{ia}\ket{_i^a} + \sum_{\bar{i}\bar{a}}c_{\bar{i}\bar{a}}\ket{_{\bar{i}}^{\bar{a}}}\right)
    \label{eqn:esmf}
\end{align}
In this work, the indices $i, j, k$ represent occupied alpha orbitals,
$a, b, c$ represent virtual alpha orbitals, and
$\bar{i}$ and $\bar{a}$ likewise represent beta orbitals.
Note that it is possible to formulate ESMF so as to also include the closed-shell
Aufbau configuration, \cite{Shea-ESMF-2018}
but we have not yet implemented the corresponding ESMP2 terms in our
$N^5$-scaling ESMP2 code, and so the ESMF reference used in this
work is as shown in Eq.\ \ref{eqn:esmf}.
Here $\hat{X}$ is an anti-Hermitian one-body operator that, when exponentiated,
produces a unitary orbital rotation that moves the linear combination of
single excitations from the HF to the ESMF orbital basis.
To find each ESMF state, we employ either the recently-introduced
generalized variational principle (GVP) \cite{Shea-GVP-2020} 
or, where possible, the more efficient ESMF self-consistent field (SCF)
approach. \cite{hardikar2020self}
The latter is not as robust as the GVP, and so we fall back to using
the GVP in cases where the SCF approach proves unstable.

\begin{figure*}
    \includegraphics[width=\textwidth]{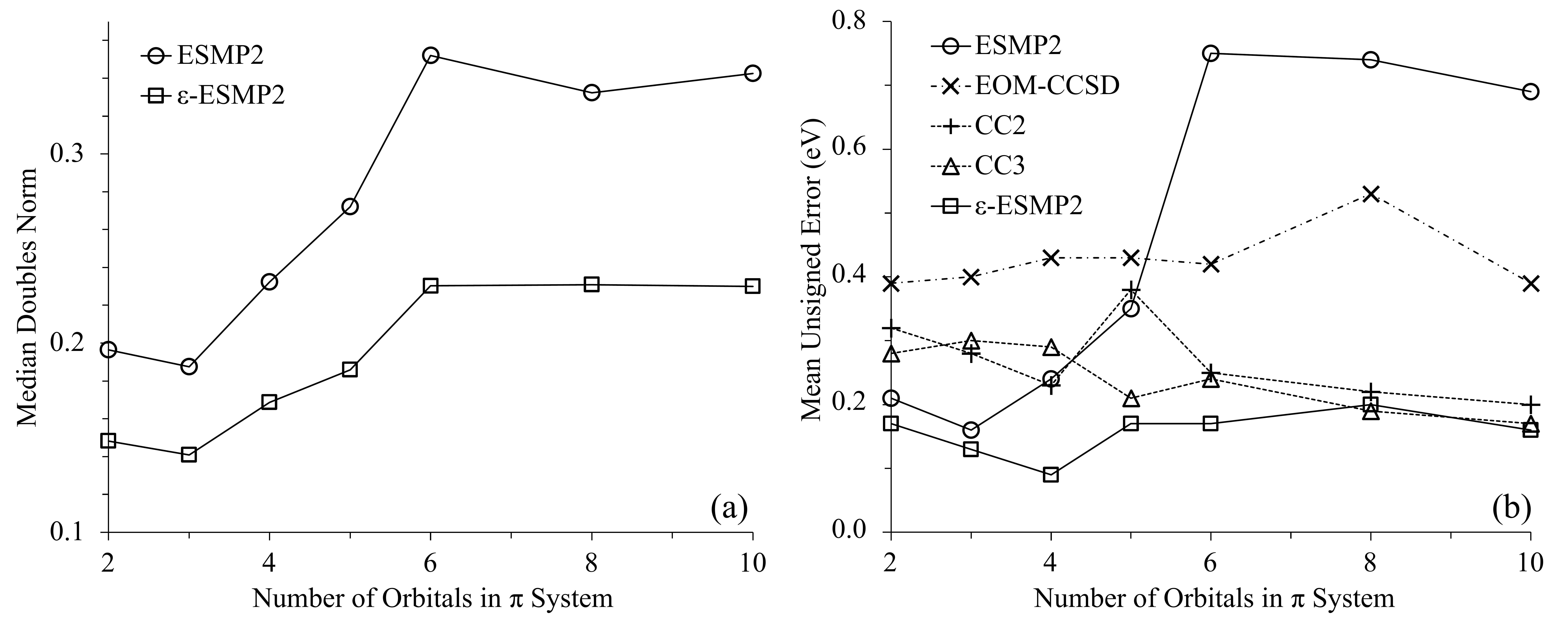}
    \caption{(a) Median $|\mathrm{T}_2|$
             doubles norms and (b) MUEs for excitation
             energies by $\pi$ system size.
             States identified by ESMP2 to have strongly doubly excited character
             and states not found by ESMF
             (red and grey rows in Table \ref{tab:all_states_with_tbe})
             are excluded.
             \label{fig:errors_pi_size}
             }
\end{figure*}

\subsection{ESMP2}

ESMP2 builds a second-order Rayleigh-Schr{\"o}dinger perturbation theory
atop ESMF in a way that parallels MP2's construction atop HF theory.
As ESMF already contains singly excited components, the initial formulation
\cite{Shea-ESMF-2018} of ESMP2 included
all doubly- and triply-excited determinants in its first order
interacting space.
This choice comes from the basic logic that if MP2 can stop at doubles
when expanding around its Aufbau reference, ESMP2 should stop at triples.
This approach led to promising accuracy in initial tests, but due to the
large number of triples, it came with an $N^7$ cost scaling.
More recently, an $N^5$ reformulation of ESMP2 has been
introduced \cite{Clune-ESMP2-2020} that includes only the most important
subset of triples by first converting the ESMF wave function into
a ``transition orbital pair'' basis that shares much in common
with the concept of a natural transition orbital basis.
\cite{martin2003ntos}
For the present study, we employ the $N^5$ theory, and refer the
reader to its original publication \cite{Clune-ESMP2-2020} for most
of its details, but let us very briefly explain the added level
shift as it has not been discussed previously.

The zeroth order Hamiltonian for ESMP2 is 
\begin{align}
    \hat{H}_0 = \hat{R} ( \hat{F} - \hat{H} ) \hat{R}
                + \hat{P} \hat{H} \hat{P}
                + \hat{Q} ( \hat{F} + \epsilon ) \hat{Q},
\end{align}
where $\hat{R}$ projects onto the ESMF state, $\hat{P}$ projects onto
the subspace containing the Aufbau and all singly-excited
configurations, and $\hat{Q}=1-\hat{P}$.
$\hat{H}$ is the full Hamiltonian, whereas $\hat{F}$ is the Fock
operator formed from the ESMF one-body density matrix.
Note that, in part, this choice of $\hat{H}_0$ is employed so as to
ensure size intensivity. \cite{Clune-ESMP2-2020}
In previous ESMP2 work, the level shift $\epsilon$ has not been
used, and all results reported as ``ESMP2'' below use $\epsilon = 0$.
By instead setting a positive
value for $\epsilon$, we can widen the zeroth order energy spacing
that separates the singles from the doubles and triples, which as
discussed above may help mitigate perturbative failures
in larger $\pi$ systems.
Due to the structure of $\hat{H}_0$, the only modification that
$\epsilon$ makes to ESMP2's working equations is to shift
the zeroth order Hamiltonian matrix's diagonal in the
amplitude equations, and so adopting a
nonzero $\epsilon$ involves a trivial algorithmic change.
As the ESMP2 excitation energy is
\begin{align}
    \Delta E = E_{\mathrm{ESMP2}} - E_{\mathrm{MP2}},
\end{align}
we also add $\epsilon$ to the denominators in the standard MP2
energy expression so as to maintain a balanced treatment
between the ground and excited state.
Of course, the value chosen for $\epsilon$ will matter,
and after some preliminary testing revealed that shifts below
0.2 $\mathrm{E_h}$ made very little difference, we ran
the full test set with the substantially larger shift value
of 0.5 $\mathrm{E_h}$ to find out what would happen with a
much more aggressive shift.
Interestingly, this resulted in substantially better
excitation energies, and so we have not attempted to
optimize $\epsilon$ any further in this study.
We show examples of individual
states' shift sensitivities in the Supplementary Material.
All results presented below that are labeled ``$\epsilon$-ESMP2''
employed $\epsilon = 0.5\ \mathrm{E_h}$.

\begin{figure*}
    \includegraphics[width=\textwidth]{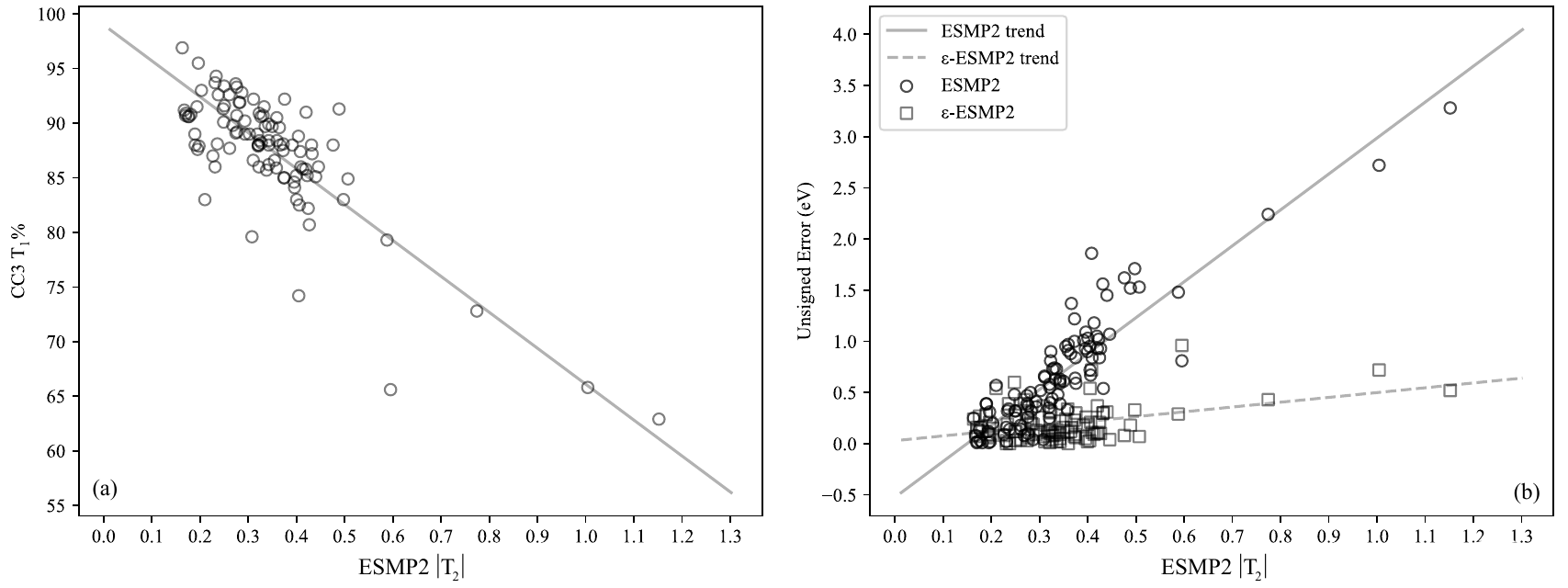}
    \caption{(a) CC3 T$_1$ percentages and
             (b) ESMP2 and $\epsilon$-ESMP2 unsigned excitation energy errors
             plotted against the ESMP2 doubles norm $|\mathrm{T}_2|$ for all states.
             The lines are linear fits to the points.
             \label{fig:vs-doubles-norm}
             }
\end{figure*}

\subsection{Amplitude Diagnostics}

Although they are not a perfect guide, \cite{karton2006w4}
amplitude diagnostics such as the T$_1$ diagnostic \cite{lee1989diagnostic}
have long been used to help predict whether ground states
are indeed weakly correlated enough for
single-reference methods to be reliable.
Might similar diagnostics offer useful information
for ESMP2?
Unlike linear response excited state methods like EOM-CCSD
in which doubly excited configurations must account for both orbital
relaxation and correlation effects, ESMP2 is built on a reference
in which mean-field orbital relaxations are already
accounted for by the MO basis.
Thus, its doubles amplitudes are more
closely related to ground state singles amplitudes:
both are singly excited relative to their reference state
and both are only expected to be present in large amounts if
the reference wave function is a poor approximation for the state.
Certainly we would not expect ESMP2 to be accurate for a state
in which any doubly excited configurations have large weights,
as this would be a violation of the assumption that we are
perturbing around the singly-excited ESMF reference.
Thus, both from their similarity to the ground state singles
at the heart of the T$_1$ diagnostic and from the perturbative
argument that they should not be large, we expect that the
ESMP2 doubles should be able to offer useful information about
the reliability of ESMP2, and possibly other theories too,
for a given excited state.

What functions of the doubles would make for good diagnostics?
In many studies involving linear-response coupled cluster theory,
the percentage of the wave function that is described by single excitations
is used as a gauge\cite{sneskov2012-WIRES}.
%and although we could adopt that or the
%percentage that is doubles here, we choose not to for reasons
%of size consistency.
While we could adopt a similar method for ESMP2, except using the percentage 
of the first order wave function coming from the doubles instead, we choose not
to as the resulting diagnostic is not size consistent. 
Instead we make use of the fact that in ESMP2
any excitation that is localized to some molecule
or molecular region (as most excitations in chemistry are) will see the
triples percentage of its wave function grow
indefinitely with system size as additional far-away molecules
are added, as the size intensivity of the theory guarantees that those
far-away molecules will simplify to MP2 descriptions,
thus adding additional triples (MP2 doubles on top of the ESMF single
excitation) components for every far-away molecule that is added.\cite{Clune-ESMP2-2020}
Therefore, the ESMP2 doubles percent
will drop to zero in the large system limit, in the same way that
the RHF determinant's percentage of the MP2 wave function goes
to zero in the large system limit.
This effect implies that the meaning of the \%T$_1$ and \%T$_2$
measures will vary with system size in ESMP2,
even when one is simply adding far-away molecules that do not
interact with the original system.
This is clearly undesirable.

In contrast, the norm of the doubles amplitudes (when working
in intermediate normalization) is unaffected by the addition of
far-away molecules,
as the size intensivity of ESMP2 guarantees that,
so long as the excitation is still on the original molecule,
the new molecules add only triples in the form of
the far-away molecules' MP2 doubles acting atop the ESMF singles.
Further, like many other diagnostics, $|$T$_2|$ is invariant to
occupied-occupied and virtual-virtual rotations.
Thus, $|$T$_2|$ offers ESMP2 a size-consistent, orbital-invariant measure
of the quality of ESMF's assumption of a purely singly-excited state.
It should therefore allow us to flag cases, like states with large doubly 
excited components, for which ESMF and ESMP2 are not appropriate.
%\textcolor{red}{
%Below, we will also consider the even simpler diagnostic of the largest
%individual doubles amplitude (also in intermediate normalization).
%Although this measure is not invariant to occupied-occupied
%and virtual-virtual rotations, it is size consistent, and our results
%contain at least one scenario in which it offers a useful complement
%to the $|$T$_2|$ diagnostic.
%}

\section{Computational Details}

Following the general considerations described by Thiel and coworkers,
\cite{Thiel-BM-MAIN}
we have employed the same ground state MP2/6-31G* geometries and 
TZVP basis set \cite{TZVP} in all calculations.
Our ESMP2 code does not currently make use of point group symmetry,
so calculations were run in C1 and manual checks were performed to ensure that
states' symmetry labels are correct.
In part to ensure the same states were being used when
comparing to existing results and in part for convenience, 
we employed the largest singles components from EOM-CCSD
calculations as the guess singles in ESMF.
We employed PySCF \cite{PYSCF} for most
EOM-CCSD calculations, while QChem \cite{qchem} 
and Molpro \cite{MOLPRO} were used for the
${2}^1$E${}_{2g}$ benzene state and the $2^1$A$_u$
and $1^1$B$_{2g}$ states of tetrazine. 
We also verified state characters by direct comparisons
of the converged ESMF and EOM-CCSD wave functions,
including visual inspection of the most relevant orbitals
for each state using Molden v2.0. \cite{molden:2017}
We further verified state character and, in particular,
the nature of doubly excited states, using
Thiel and coworkers' active spaces \cite{Thiel-BM-MAIN}
and Molpro's implementation of state-averaged CASSCF.
Detailed information on these various comparisons
can be found in the Supplementary Material. 
Note that in some of our comparisons below, we have
excluded states not found by ESMF or that are
flagged by ESMP2 as having large amounts of doubly
excited character, as these either cannot be compared
or are not appropriate for treatment
by any of the single-reference methods.
We have verified that crunching the numbers with
these states included makes little difference,
as discussed in the next section and as seen
in the additional tables in the SI.

%\begin{table}
%\caption{
%  \label{tab:excluded}
%  \textcolor{red}{
%  States excluded from the analysis in Table XYZ due to the presence of doubly excited character.}
%}
%%\begin{tabular}{l l l}
%\begin{tabular}{l l}
%\hline\hline
%\hspace{1mm} Molecule \rule{0pt}{3.8mm} &
%\hspace{1mm} State 
%%\hspace{1mm} Reason
%\\[2pt]
%\hline
%\hspace{1mm} butadiene                      & \hspace{1mm} $2^1A_g$    \\[2pt] %& \hspace{1mm} doubly excited character \\[2pt]
%\hspace{1mm} hexatriene                     & \hspace{1mm} $2^1A_g$   \\[2pt] % & \hspace{1mm} doubly excited character \\[2pt]
%\hspace{1mm} octatetraene                   & \hspace{1mm} $2^1A_g$    \\[2pt] %& \hspace{1mm} doubly excited character \\[2pt]
%\hspace{1mm} benzoquinone                   & \hspace{1mm} $1^1B_{3u}$ \\[2pt] %& \hspace{1mm} doubly excited character \\[2pt]
%%\hspace{1mm} cyclopentadiene                & \hspace{1mm} $2^1A_1$    & \hspace{1mm} doubly excited character \\[2pt]
%\hspace{1mm} benzene                        & \hspace{1mm} $2^1E_{2g}$ \\[2pt] %& \hspace{1mm} doubly excited character \\[2pt]
%\hline\hline
%%\vspace{2mm}
%\end{tabular}
%\end{table}

\section{Results}

\subsection{Overview}

Table \ref{tab:all_states_with_tbe}
shows our results on the 103 singlet states
that have CC results and TBEs in the Thiel benchmark, \cite{Thiel-BM-MAIN}
with the ESMP2 and CC methods' accuracies
summarized in Figure \ref{fig:errors_pi_size}
and Table \ref{tab:wfn_red_gray}.
Orbital-optimized ESMF stationary points were
successfully located for 100 of these 103 states, which, while
not perfect, represents the clearest evidence to date that
ESMF energy stationary points can be expected to exist
for the vast majority of low-lying singlet excited states
in single-reference molecules.
Six of the states showed especially large ESMP2 doubles
norms with $|\mathrm{T}_2|>0.5$, and
five of these six likewise had CC3 T1 percentages below 80,
indicating that ESMP2's $|\mathrm{T}_2|$ can indeed help predict
states with challenging amounts of doubly excited character.
As seen in Figure \ref{fig:vs-doubles-norm},
ESMP2's $|\mathrm{T}_2|$ also shows the expected correlations
with both ESMP2 excitation energy errors and the CC3 T1 percentage
across a wider range of $|\mathrm{T}_2|$ values.
With regards to excluding states from some comparisons,
we note that the exclusion of the three missing ESMF states from
the statistics, for example in Table \ref{tab:wfn_red_gray},
changes the overall mean unsigned error (MUE) by 0.01 eV
or less for CC2, EOM-CCSD, and CC3.
Excluding the states with $|\mathrm{T}_2|>0.5$
improves the overall MUEs of CC2, EOM-CCSD, CC3,
and $\epsilon$-ESMP2 by just
0.04, 0.03, 0.0, and 0.02 eV, respectively, and
so does not affect the ordering of their overall accuracy.
%Thus, the exclusions do not alter any conclusions, while at the
%same time focusing the comparisons onto the states where we
%have ESMP2 data and where the use of single-reference
%methods is most appropriate.
As there were only three states out of about 100 that ESMF
did not find, we do not expect their absence from the ESMP2 statistics
to alter any of the broad conclusions drawn from this study.
The Supplementary Material has additional tables in which
fewer states are excluded.

As seen in Figure \ref{fig:errors_pi_size},
unregularized ESMP2's accuracy for excitation energies
in singly-excited states depends strongly on the size of a molecule's
$\pi$ system, while $\epsilon$-ESMP2
is insensitive to $\pi$ system size and highly accurate.
The degradation of ESMP2's accuracy with increasing $\pi$
system size closely follows the rise of its $|\mathrm{T}_2|$ doubles norm,
indicating that the poor accuracy in molecules with larger $\pi$ systems
is indeed related to a perturbative failure
born of small zeroth order energy spacings between the reference and the
lowest-lying doubles.
With its level shift suppressing the spurious growth of large doubles
contributions, $\epsilon$-ESMP2 is significantly more reliable,
displaying an accuracy that is as good or better than the other
single-reference methods at all $\pi$ system sizes.
Among the methods compared, only the multi-reference CASPT2
approach using Thiel's active spaces
offers better overall accuracy on this test set than $\epsilon$-ESMP2.
With an $N^5$ scaling and no need to choose an active space, these
results suggest that $\epsilon$-ESMP2 has much to offer in modeling
singly excited states, while ESMP2 can act as a relatively affordable
detector of doubly excited character.

%The molecules studied here have $\pi$ system sizes
%(defined as the number of 2p orbitals that participate
%in the $\pi$ system) ranging from 2 to 10, and, as seen in
%Table XYZ, the accuracy of unshifted ESMP2 depends strongly on
%this property.
%Essentially, larger $\pi$ systems lead to a more compressed
%orbital energy spectrum, which in turn creates more near-degeneracies
%between different orbitals and excitations.
%These near-degeneracies are anathema to MP2-style
%perturbation theories and are primary motivators for
%regularization approaches like level shifts.
%Interestingly, adding a 0.5 E$_h$ level shift to the
%diagonal of ESMP2's 0$^{\mathrm{th}}$ order Hamiltonian
%leads to greatly improved results and a substantially
%smaller sensitivity to $\pi$ system size.
%On this test set, only CASPT2 offers a lower MAD than this
%level-shifted ESMP2.
%We note that the CC2, CC3, EOM-CCSD, and CASPT2
%MADs and STDEVs in Table XYZ change by less than ??? eV
%if we include the three states that lack ESMF solutions,
%and so the exclusion of these states does not
%change any of our conclusions.
%Before we move on to analyzing results in more detail,
%the reader is encouraged to peruse the state-by-state
%excitation energies of all 93 states in Table
%???.

\subsection{Amplitude Diagnostics}

As seen in Figure \ref{fig:vs-doubles-norm}, the
ESMP2 $|\mathrm{T}_2|$ values tend to increase
as the CC3 \%T$_1$ values decrease, in line
with expectation.
With one exception, the most worrying CC3 \%T$_1$ values
(those significantly below 80\%) all correspond to
ESMP2 $|\mathrm{T}_2|$ values above 0.5.
The exception is the 1$^1$B$_{\mathrm{3u}}$ state
of benzoquinone, which has a CC3 \%T$_1$ of 75.2\%
but an ESMP2 doubles norm of just 0.4.  
Interestingly, both CC2 and CC3 are reasonably
accurate for this state despite the low \%T$_1$ value,
although EOM-CCSD and $\epsilon$-ESMP2 are not.
This exception makes it tempting to recommend that
states with $|\mathrm{T}_2| > 0.4$ be considered ``hard''
for $\epsilon$-ESMP2, but Figure \ref{fig:vs-doubles-norm}
also makes clear that there are many states with
doubles norms this large that $\epsilon$-ESMP2 is
quite accurate for, and a couple with lower
doubles norms where $\epsilon$-ESMP2 struggles.
So we see 0.5 as a better rough threshold for when
to firmly set $\epsilon$-ESMP2 and other single-reference
methods aside and reach for multi-reference approaches.
For ESMP2, large energy errors clearly start much
earlier, and it would be difficult to recommend relying
on it for any state where $|\mathrm{T}_2| > 0.3$.

%couldn't get include to work, error was related to nesting, but I can't figure out why
\begin{longtable*}{l c l| c c c c c c | c c c}
\caption{
Singlet excitation energies in eV.
TBEs and results for CASPT2, CC2, EOM-CCSD,
and CC3 are from the original Thiel benchmark, \cite{Thiel-BM-MAIN}
except for CC3 results on cytosine, thymine, uracil, and adenine,
which are from K{\'a}nn{\'a}r and Szalay.
\cite{kannar2014-nucleobases}
CASPT2 ``a'' refers to Roos's results,
while CASPT2 ``b'' refers to Thiel's.
States where no ESMF solution was found are highlighted in gray,
those with known Rydberg character are in blue, and those
in which ESMP2's $|\mathrm{T}_2|$ was above 0.5 are in red.
\label{tab:all_states_with_tbe}
}
\\
\hline\hline
\hspace{2mm}  Molecule &
\hspace{0mm} $\pi$ &
\hspace{2mm} State & 
\hspace{2mm} TBE & 
\hspace{2mm} CASPT2 & 
\hspace{2mm} CASPT2 & 
\hspace{2mm} CC2 & 
\hspace{2mm} CCSD & 
\hspace{2mm} CC3 & 
\hspace{2mm} ESMP2 & 
\hspace{2mm} $\epsilon$-ESMP2 &
\hspace{2mm} ESMP2
\\[2pt] 
\hspace{2mm} & size & & & a & b & & & \hspace{2mm}  (\%T\textsubscript{1}) & & & \hspace{2mm}
$|\mathrm{T}_2|$
\\[2pt] 
\hline 
\hspace{0.0mm} formaldehyde & 2 & $1^1A_{2}$ & 3.88 & 3.91 & 3.98 & 4.09 & 3.97 & 3.95  (91.2) & 3.96 & 3.81 & 0.17\\ 
\hspace{0.0mm} & & $1^1B_{1}$ & 9.10 & 9.09 & 9.14 & 9.35 & 9.26 & 9.18 (90.9) & 9.12 & 8.98 & 0.17\\
%\rowcolor{DeepSkyBlue1!25}\hspace{0.0mm} & & $2^1A_{1}$ & 9.30 & 9.77 & 9.31 & 10.34 & 10.54 & 10.45 (91.3) & 9.78 & 9.90 & 0.25\\ &&&&&&&&&&& \\[-0.15cm] % old way of coloring the table and separating molecules
\hspace{0.0mm} & & \cellcolor{DeepSkyBlue1!25}$2^1A_{1}$ & \cellcolor{DeepSkyBlue1!25}9.30 & \cellcolor{DeepSkyBlue1!25}9.77 & \cellcolor{DeepSkyBlue1!25}9.31 & \cellcolor{DeepSkyBlue1!25}10.34 & \cellcolor{DeepSkyBlue1!25}10.54 & \cellcolor{DeepSkyBlue1!25}10.45 (91.3) & \cellcolor{DeepSkyBlue1!25}9.78 & \cellcolor{DeepSkyBlue1!25}9.90 & \cellcolor{DeepSkyBlue1!25}0.25\\ 
&&&&&&&&&&& \\[-0.15cm]
\hspace{0.0mm} acetone & 2 & $1^1A_{2}$ & 4.40 & 4.18 & 4.42 & 4.52 & 4.43 & 4.40 (90.8) & 4.39 & 4.28 & 0.18\\ 
\hspace{0.0mm} & & $1^1B_{1}$ & 9.10 & 9.10 & 9.27 & 9.29 & 9.26 & 9.17 (91.5) & 9.22 & 9.16 & 0.19\\ 
\hspace{0.0mm} & & \cellcolor{DeepSkyBlue1!25}$2^1A_{1}$ & \cellcolor{DeepSkyBlue1!25}9.40 & \cellcolor{DeepSkyBlue1!25}9.16 & \cellcolor{DeepSkyBlue1!25}9.31 & \cellcolor{DeepSkyBlue1!25}9.74 & \cellcolor{DeepSkyBlue1!25}9.87 & \cellcolor{DeepSkyBlue1!25}9.65 (90.1) & \cellcolor{DeepSkyBlue1!25}9.08 & \cellcolor{DeepSkyBlue1!25}9.28 & \cellcolor{DeepSkyBlue1!25}0.25\\ &&&&&&&&&&& \\[-0.15cm]
\hspace{0.0mm} benzoquinone & 8 & $1^1A_{u}$ & 2.80 & 2.50 & 2.80 & 2.92 & 3.19 & 2.85 (83.0) & 1.77 & 2.61 & 0.40\\ 
\hspace{0.0mm} & & $1^1B_{1g}$ & 2.78 & 2.50 & 2.78 & 2.81 & 3.07 & 2.75 (84.1) & 1.69 & 2.52 & 0.40\\ 
\hspace{0.0mm} & & $1^1B_{3g}$ & 4.25 & 4.19 & 4.25 & 4.69 & 4.93 & 4.59 (87.9) & 3.67 & 4.12 & 0.32\\ 
\hspace{0.0mm} & & $1^1B_{1u}$ & 5.29 & 5.15 & 5.29 & 5.59 & 5.89 & 5.62 (88.4) & 4.70 & 5.31 & 0.34\\ 
\hspace{0.0mm} & & $1^1B_{3u}$ & 5.60 & 5.15 & 5.60 & 5.69 & 6.55 & 5.82 (75.2) & 4.88 & 6.14 & 0.40\\ 
\hspace{0.0mm} & & $2^1B_{3g}$ & 6.98 & 6.34 & 6.98 & 7.36 & 7.62 & 7.27 (88.8) & 6.03 & 7.01 & 0.40\\ &&&&&&&&&&& \\[-0.15cm]
\hspace{0.0mm} formamide & 3 & $1^1A^{\prime\prime}$ & 5.63 & 5.61 & 5.63 & 5.76 & 5.66 & 5.65 (90.7) & 5.62 & 5.47 & 0.17\\ 
\hspace{0.0mm} & & $2^1A'$ & 7.44 & 7.41 & 7.44 & 8.15 & 8.52 & 8.27 (87.9) & 7.42 & 7.52 & 0.20\\ &&&&&&&&&&& \\[-0.15cm]
\hspace{0.0mm} acetamide & 3 & $1^1A^{\prime\prime}$ & 5.80 & 5.54 & 5.80 & 5.77 & 5.71 & 5.69 (90.6) & 5.66 & 5.53 & 0.18\\ 
\hspace{0.0mm} & & $2^1A'$ & 7.27 & 7.21 & 7.27 & 7.66 & 7.85 & 7.67 (89.1) & 6.88 & 7.30 & 0.27\\ &&&&&&&&&&& \\[-0.15cm]
\hspace{0.0mm} propanamide & 3 & $1^1A^{\prime\prime}$ & 5.72 & 5.48 & 5.72 & 5.78 & 5.74 & 5.72 (90.6) & 5.69 & 5.56 & 0.18\\ 
\hspace{0.0mm} & & $2^1A'$ & 7.20 & 7.28 & 7.20 & 7.56 & 7.80 & 7.62 (89.2) & 6.83 & 7.26 & 0.28\\ &&&&&&&&&&& \\[-0.15cm]
\hspace{0.0mm} ethene & 2 & \cellcolor{DeepSkyBlue1!25}$1^1B_{1u}$ & \cellcolor{DeepSkyBlue1!25}7.80 & \cellcolor{DeepSkyBlue1!25}7.98 & \cellcolor{DeepSkyBlue1!25}8.62 & \cellcolor{DeepSkyBlue1!25}8.40 & \cellcolor{DeepSkyBlue1!25}8.51 & \cellcolor{DeepSkyBlue1!25}8.37 (96.9) & \cellcolor{DeepSkyBlue1!25}8.05 & \cellcolor{DeepSkyBlue1!25}8.04 & \cellcolor{DeepSkyBlue1!25}0.16\\ &&&&&&&&&&& \\[-0.15cm]
\hspace{0.0mm} butadiene & 4 & $1^1B_{u}$ & 6.18 & 6.23 & 6.47 & 6.49 & 6.72 & 6.58 (93.7) & 6.02 & 6.18 & 0.23\\ 
\hspace{0.0mm} & & \cellcolor{Firebrick2!25}$2^1A_{g}$ & \cellcolor{Firebrick2!25}6.55 & \cellcolor{Firebrick2!25}6.27 & \cellcolor{Firebrick2!25}6.83 & \cellcolor{Firebrick2!25}7.63 & \cellcolor{Firebrick2!25}7.42 & \cellcolor{Firebrick2!25}6.77 (72.8) & \cellcolor{Firebrick2!25}4.31 & \cellcolor{Firebrick2!25}6.98 & \cellcolor{Firebrick2!25}0.77\\ &&&&&&&&&&& \\[-0.15cm]
\hspace{0.0mm} hexatriene & 6 & $1^1B_{u}$ & 5.10 & 5.01 & 5.31 & 5.41 & 5.72 & 5.58 (92.6) & 4.92 & 5.14 & 0.26\\ 
\hspace{0.0mm} & & \cellcolor{Firebrick2!25}$2^1A_{g}$ & \cellcolor{Firebrick2!25}5.09 & \cellcolor{Firebrick2!25}5.20 & \cellcolor{Firebrick2!25}5.42 & \cellcolor{Firebrick2!25}6.67 & \cellcolor{Firebrick2!25}6.61 & \cellcolor{Firebrick2!25}5.72 (65.8) & \cellcolor{Firebrick2!25}2.37 & \cellcolor{Firebrick2!25}5.81 & \cellcolor{Firebrick2!25}1.00\\ &&&&&&&&&&& \\[-0.15cm]
\hspace{0.0mm} octatetraene & 8 & \cellcolor{Firebrick2!25}$2^1A_{g}$ & \cellcolor{Firebrick2!25}4.47 & \cellcolor{Firebrick2!25}4.38 & \cellcolor{Firebrick2!25}4.64 & \cellcolor{Firebrick2!25}5.87 & \cellcolor{Firebrick2!25}5.99 & \cellcolor{Firebrick2!25}4.97 (62.9) & \cellcolor{Firebrick2!25}1.19 & \cellcolor{Firebrick2!25}4.99 & \cellcolor{Firebrick2!25}1.15 \\ 
\hspace{0.0mm} & & $1^1B_{u}$ & 4.66 & 4.42 & 4.70 & 4.72 & 5.07 & 4.94 (91.9) & 4.16 & 4.43 & 0.28\\ &&&&&&&&&&& \\[-0.15cm]
\hspace{0.0mm} cyclopropene & 2 & $1^1B_{1}$ & 6.76 & 6.36 & 6.76 & 6.96 & 6.96 & 6.90 (93.0) & 6.56 & 6.61 & 0.20\\ 
\hspace{0.0mm} & & \cellcolor{DeepSkyBlue1!25}$1^1B_{2}$ & \cellcolor{DeepSkyBlue1!25}7.06 & \cellcolor{DeepSkyBlue1!25}7.45 & \cellcolor{DeepSkyBlue1!25}7.06 & \cellcolor{DeepSkyBlue1!25}7.17 & \cellcolor{DeepSkyBlue1!25}7.24 & \cellcolor{DeepSkyBlue1!25}7.10 (95.5) & \cellcolor{DeepSkyBlue1!25}6.75 & \cellcolor{DeepSkyBlue1!25}6.85 & \cellcolor{DeepSkyBlue1!25}0.20\\ &&&&&&&&&&& \\[-0.15cm]
\hspace{0.0mm} cyclopentadiene & 4 & $1^1B_{2}$ & 5.55 & 5.27 & 5.51 & 5.69 & 5.87 & 5.73 (94.3) & 5.23 & 5.36 & 0.23\\ 
\hspace{0.0mm} & & \cellcolor{Firebrick2!25}$2^1A_{1}$ & \cellcolor{Firebrick2!25}6.31 & \cellcolor{Firebrick2!25}6.31 & \cellcolor{Firebrick2!25}6.31 & \cellcolor{Firebrick2!25}7.05 & \cellcolor{Firebrick2!25}7.05 & \cellcolor{Firebrick2!25}6.61 (79.3) & \cellcolor{Firebrick2!25}4.83 & \cellcolor{Firebrick2!25}6.60 & \cellcolor{Firebrick2!25}0.60\\ &&&&&&&&&&& \\[-0.15cm]
\hspace{0.0mm} norbornadiene & 2 & $1^1A_{2}$ & 5.34 & 5.28 & 5.34 & 5.57 & 5.80 & 5.64 (93.4) & 5.09 & 5.31 & 0.25\\ 
\hspace{0.0mm} & & \cellcolor{DeepSkyBlue1!25}$1^1B_{2}$ & \cellcolor{DeepSkyBlue1!25}6.11 & \cellcolor{DeepSkyBlue1!25}6.20 & \cellcolor{DeepSkyBlue1!25}6.11 & \cellcolor{DeepSkyBlue1!25}6.37 & \cellcolor{DeepSkyBlue1!25}6.69 & \cellcolor{DeepSkyBlue1!25}6.49 (91.1) & \cellcolor{DeepSkyBlue1!25}5.79 & \cellcolor{DeepSkyBlue1!25}6.31 & \cellcolor{DeepSkyBlue1!25}0.28 \\ &&&&&&&&&&& \\[-0.15cm]
\hspace{0.0mm} benzene & 6 & $1^1B_{2u}$ & 5.08 & 4.84 & 5.05 & 5.27 & 5.19 & 5.07 (85.8) & 4.03 & 4.98 & 0.42\\ 
\hspace{0.0mm} & & $1^1B_{1u}$ & 6.54 & 6.30 & 6.45 & 6.68 & 6.74 & 6.68 (93.6) & 6.07 & 6.24 & 0.27\\ 
\hspace{0.0mm} & & $1^1E_{1u}$ & 7.13 & 7.03 & 7.07 & 7.44 & 7.65 & 7.45 (92.2) & 6.48 & 7.11 & 0.31\\ 
\hspace{0.0mm} & & \cellcolor{Firebrick2!25}$2^1E_{2g}$ & \cellcolor{Firebrick2!25}8.41 & \cellcolor{Firebrick2!25}7.90 & \cellcolor{Firebrick2!25}8.21 & \cellcolor{Firebrick2!25}9.03 & \cellcolor{Firebrick2!25}9.21 & \cellcolor{Firebrick2!25}8.43 (65.6) & \cellcolor{Firebrick2!25}7.60 & \cellcolor{Firebrick2!25}9.37 & \cellcolor{Firebrick2!25}0.60\\ &&&&&&&&&&& \\[-0.15cm]
\hspace{0.0mm} naphthalene & 10 & $1^1B_{3u}$ & 4.24 & 4.03 & 4.24 & 4.45 & 4.41 & 4.27 (85.2) & 3.22 & 4.13 & 0.42 \\ 
\hspace{0.0mm} & & $1^1B_{2u}$ & 4.77 & 4.56 & 4.77 & 4.96 & 5.21 & 5.03 (90.6) & 4.33 & 4.73 & 0.33\\ 
\hspace{0.0mm} & & $2^1A_{g}$ & 5.90 & 5.39 & 5.90 & 6.22 & 6.23 & 5.98 (82.2) & 5.06 & 6.11 & 0.42\\ 
\hspace{0.0mm} & & $1^1B_{1g}$ & 6.00 & 5.53 & 6.00 & 6.21 & 6.53 & 6.07 (79.6) & 5.96 & 6.37 & 0.31\\ 
\hspace{0.0mm} & & $2^1B_{3u}$ & 6.07 & 5.54 & 6.07 & 6.25 & 6.55 & 6.33 (90.7) & 5.33 & 5.99 & 0.33\\ 
\hspace{0.0mm} & & $2^1B_{1g}$ & 6.48 & 5.87 & 6.48 & 6.82 & 6.97 & 6.79 (91.3) & 4.96 & 6.30 & 0.49\\ 
\hspace{0.0mm} & & $2^1B_{2u}$ & 6.33 & 5.93 & 6.33 & 6.57 & 6.77 & 6.57 (90.5) & 5.42 & 6.22 & 0.36\\ 
\hspace{0.0mm} & & \cellcolor{lightgray}$3^1A_{g}$ & \cellcolor{lightgray}6.71 & \cellcolor{lightgray}6.04 & \cellcolor{lightgray}6.71 & \cellcolor{lightgray}7.34 & \cellcolor{lightgray}7.77 & \cellcolor{lightgray}6.90 (70.0) & \cellcolor{lightgray}N/A & \cellcolor{lightgray}N/A & \cellcolor{lightgray}N/A \\ &&&&&&&&&&& \\[-0.15cm]
\hspace{0.0mm} furan & 5 & $1^1B_{2}$ & 6.32 & 6.04 & 6.43 & 6.43 & 6.80 & 6.60 (92.6) & 6.18 & 6.32 & 0.24\\ 
\hspace{0.0mm} & & \cellcolor{Firebrick2!25}$2^1A_{1}$ & \cellcolor{Firebrick2!25}6.57 & \cellcolor{Firebrick2!25}6.16 & \cellcolor{Firebrick2!25}6.52 & \cellcolor{Firebrick2!25}6.87 & \cellcolor{Firebrick2!25}6.89 & \cellcolor{Firebrick2!25}6.62 (84.9) & \cellcolor{Firebrick2!25}5.04 & \cellcolor{Firebrick2!25}6.50 & \cellcolor{Firebrick2!25}0.51\\ 
\hspace{0.0mm} & & \cellcolor{DeepSkyBlue1!25}$3^1A_{1}$ & \cellcolor{DeepSkyBlue1!25}8.13 & \cellcolor{DeepSkyBlue1!25}7.66 & \cellcolor{DeepSkyBlue1!25}8.22 & \cellcolor{DeepSkyBlue1!25}8.83 & \cellcolor{DeepSkyBlue1!25}8.83 & \cellcolor{DeepSkyBlue1!25}8.53 (90.7) & \cellcolor{DeepSkyBlue1!25}7.87 & \cellcolor{DeepSkyBlue1!25}8.38 & \cellcolor{DeepSkyBlue1!25}0.28\\ &&&&&&&&&&& \\[-0.15cm] 
\hspace{0.0mm} pyrrole & 5 & $2^1A_{1}$ & 6.37 & 5.92 & 6.31 & 6.61 & 6.61 & 6.40 (86.0) & 5.30 & 6.41 & 0.45\\ 
\hspace{0.0mm} & & \cellcolor{DeepSkyBlue1!25}$1^1B_{2}$ & \cellcolor{DeepSkyBlue1!25}6.57 & \cellcolor{DeepSkyBlue1!25}6.00 & \cellcolor{DeepSkyBlue1!25}6.33 & \cellcolor{DeepSkyBlue1!25}6.83 & \cellcolor{DeepSkyBlue1!25}6.87 & \cellcolor{DeepSkyBlue1!25}6.71 (91.6) & \cellcolor{DeepSkyBlue1!25}6.25 & \cellcolor{DeepSkyBlue1!25}6.45 & \cellcolor{DeepSkyBlue1!25}0.25\\ 
\hspace{0.0mm} & & \cellcolor{DeepSkyBlue1!25}$3^1A_{1}$ & \cellcolor{DeepSkyBlue1!25}7.91 & \cellcolor{DeepSkyBlue1!25}7.46 & \cellcolor{DeepSkyBlue1!25}8.17 & \cellcolor{DeepSkyBlue1!25}8.44 & \cellcolor{DeepSkyBlue1!25}8.44 & \cellcolor{DeepSkyBlue1!25}8.17 (90.2) & \cellcolor{DeepSkyBlue1!25}7.50 & \cellcolor{DeepSkyBlue1!25}8.04 & \cellcolor{DeepSkyBlue1!25}0.29\\ &&&&&&&&&&& \\[-0.15cm]
\hspace{0.0mm} imidazole & 5 & $1^1A^{\prime\prime}$ & 6.81 & 6.52 & 6.81 & 6.86 & 7.01 & 6.82 (87.6) & 6.80 & 6.72 & 0.19\\ 
\hspace{0.0mm} & & $2^1A'$ & 6.19 & 6.72 & 6.19 & 6.73 & 6.80 & 6.58 (87.2) & 5.65 & 6.49 & 0.43\\ 
\hspace{0.0mm} & & $3^1A'$ & 6.93 & 7.15 & 6.93 & 7.28 & 7.27 & 7.10 (89.8) & 7.01 & 7.36 & 0.27\\ &&&&&&&&&&& \\[-0.15cm]
\hspace{0.0mm} pyridine & 6 & $1^1B_{2}$ & 4.85 & 4.84 & 5.02 & 5.32 & 5.27 & 5.15 (85.9) & 4.52 & 5.19 & 0.36\\ 
\hspace{0.0mm} & & $1^1B_{1}$ & 4.59 & 4.91 & 5.14 & 5.12 & 5.25 & 5.05 (88.1) & 4.93 & 4.98 & 0.24\\ 
\hspace{0.0mm} & & $2^1A_{2}$ & 5.11 & 5.17 & 5.47 & 5.39 & 5.73 & 5.50 (87.7) & 5.25 & 5.50 & 0.26\\ 
\hspace{0.0mm} & & $2^1A_{1}$ & 6.26 & 6.42 & 6.39 & 6.88 & 6.94 & 6.85 (92.8) & 6.22 & 6.45 & 0.29\\ 
\hspace{0.0mm} & & $3^1A_{1}$ & 7.18 & 7.23 & 7.46 & 7.72 & 7.94 & 7.70 (91.5) & 6.55 & 7.29 & 0.33\\ 
\hspace{0.0mm} & & $2^1B_{2}$ & 7.27 & 7.48 & 7.29 & 7.61 & 7.81 & 7.59 (89.7) & 6.54 & 7.25 & 0.34\\ &&&&&&&&&&& \\[-0.15cm]
\hspace{0.0mm} pyrazine & 6 & $1^1B_{3u}$ & 3.95 & 3.63 & 4.12 & 4.26 & 4.42 & 4.24 (89.9) & 3.34 & 4.00 & 0.34\\ 
\hspace{0.0mm} & & $1^1A_{u}$ & 4.81 & 4.52 & 4.70 & 4.95 & 5.29 & 5.05 (88.4) & 3.84 & 4.81 & 0.36\\ 
\hspace{0.0mm} & & $1^1B_{2u}$ & 4.64 & 4.75 & 4.85 & 5.13 & 5.14 & 5.02 (86.2) & 4.25 & 4.86 & 0.34\\ 
\hspace{0.0mm} & & $1^1B_{2g}$ & 5.56 & 5.17 & 5.68 & 5.92 & 6.02 & 5.74 (85.0) & 4.92 & 5.65 & 0.37\\ 
\hspace{0.0mm} & & $1^1B_{1g}$ & 6.60 & 6.13 & 6.41 & 6.70 & 7.13 & 6.75 (85.8) & 5.42 & 6.69 & 0.41\\ 
\hspace{0.0mm} & & $1^1B_{1u}$ & 6.58 & 6.70 & 6.89 & 7.10 & 7.18 & 7.07 (93.3) & 6.48 & 6.66 & 0.28\\ 
\hspace{0.0mm} & & $2^1B_{1u}$ & 7.72 & 7.57 & 7.79 & 8.13 & 8.34 & 8.06 (90.9) & 6.91 & 7.62 & 0.32\\ 
\hspace{0.0mm} & & $2^1B_{2u}$ & 7.60 & 7.70 & 7.65 & 8.07 & 8.29 & 8.05 (89.7) & 6.99 & 7.76 & 0.35\\ &&&&&&&&&&& \\[-0.15cm]
\hspace{0.0mm} pyrimidine & 6 & $1^1B_{1}$ & 4.55 & 3.81 & 4.44 & 4.49 & 4.70 & 4.50 (88.4) & 3.65 & 4.24 & 0.32\\ 
\hspace{0.0mm} & & $1^1A_{2}$ & 4.91 & 4.12 & 4.81 & 4.84 & 5.12 & 4.93 (88.2) & 4.18 & 4.83 & 0.33\\ 
\hspace{0.0mm} & & $1^1B_{2}$ & 5.44 & 4.93 & 5.24 & 5.51 & 5.49 & 5.36 (85.7) & 4.96 & 5.55 & 0.34\\ 
\hspace{0.0mm} & & $2^1A_{1}$ & 6.95 & 6.72 & 6.64 & 7.12 & 7.17 & 7.06 (92.2) & 6.36 & 7.25 & 0.38\\ &&&&&&&&&&& \\[-0.15cm]
\hspace{0.0mm} pyridazine & 6 & $1^1B_{1}$ & 3.78 & 3.48 & 3.78 & 3.90 & 4.11 & 3.92 (89.0) & 3.26 & 3.67 & 0.30\\ 
\hspace{0.0mm} & & $1^1A_{2}$ & 4.32 & 3.66 & 4.32 & 4.40 & 4.76 & 4.49 (86.6) & 3.66 & 4.25 & 0.31\\ 
\hspace{0.0mm} & & $2^1A_{1}$ & 5.18 & 4.86 & 5.18 & 5.37 & 5.35 & 5.22 (85.2) & 4.28 & 5.16 & 0.40\\ 
\hspace{0.0mm} & & $2^1A_{2}$ & 5.77 & 5.09 & 5.77 & 5.81 & 6.00 & 5.74 (86.6) & 4.82 & 5.53 & 0.35\\ &&&&&&&&&&& \\[-0.15cm]
\hspace{0.0mm} triazine & 6 & $1^1A_{1}^{\prime\prime}$ & 4.60 & 3.90 & 4.60 & 4.70 & 4.96 & 4.78 (88.0) & 3.23 & 4.44 & 0.37\\ 
\hspace{0.0mm} & & $1^1A_{2}^{\prime\prime}$ & 4.66 & 4.08 & 4.68 & 4.80 & 4.98 & 4.76 (88.0) & 3.65 & 4.48 & 0.39\\ 
\hspace{0.0mm} & & $1^1E^{\prime\prime}$ & 4.71 & 4.36 & 4.71 & 4.77 & 5.01 & 4.81 (88.1) & 3.49 & 4.48 & 0.37\\ 
\hspace{0.0mm} & & $1^1A_2^\prime$ & 5.79 & 5.33 & 5.79 & 5.82 & 5.84 & 5.71 (85.1) & 4.34 & 5.48 & 0.44\\ &&&&&&&&&&& \\[-0.15cm]
\hspace{0.0mm} tetrazine & 6 & $1^1B_{3u}$ & 2.24 & 1.96 & 2.24 & 2.47 & 2.71 & 2.53 (89.6) & 1.36 & 2.12 & 0.36\\ 
\hspace{0.0mm} & & $1^1A_{u}$ & 3.48 & 3.06 & 3.48 & 3.67 & 4.07 & 3.79 (87.5) & 2.48 & 3.55 & 0.37\\ 
\hspace{0.0mm} & & $1^1B_{1g}$ & 4.73 & 4.51 & 4.73 & 5.10 & 5.32 & 4.97 (82.5) & 4.06 & 4.94 & 0.41\\ 
\hspace{0.0mm} & & $1^1B_{2u}$ & 4.91 & 4.89 & 4.91 & 5.20 & 5.27 & 5.12 (84.6) & 3.98 & 4.86 & 0.40\\ 
\hspace{0.0mm} & & $2^1A_{u}$ & 5.47 & 5.28 & 5.47 & 5.50 & 5.70 & 5.46 (87.4) & 3.61 & 4.75 & 0.41\\ 
\hspace{0.0mm} & & $1^1B_{2g}$ & 5.18 & 5.05 & 5.18 & 5.53 & 5.70 & 5.34 (80.7) & 4.25 & 5.28 & 0.43\\ &&&&&&&&&&& \\[-0.15cm]
\hspace{0.0mm} cytosine & 8 & \cellcolor{lightgray}$2^1A'$ & \cellcolor{lightgray}4.66 & \cellcolor{lightgray}4.39 & \cellcolor{lightgray}4.68 & \cellcolor{lightgray}4.80 & \cellcolor{lightgray}4.98 & \cellcolor{lightgray}4.72 (86)& \cellcolor{lightgray}N/A & \cellcolor{lightgray}N/A & \cellcolor{lightgray}N/A\\
\hspace{0.0mm} & & \cellcolor{DeepSkyBlue1!25}$1^1A^{\prime\prime}$ & \cellcolor{DeepSkyBlue1!25}4.87 & \cellcolor{DeepSkyBlue1!25}5.00 & \cellcolor{DeepSkyBlue1!25}5.12 & \cellcolor{DeepSkyBlue1!25}5.13 & \cellcolor{DeepSkyBlue1!25}5.45 & \cellcolor{DeepSkyBlue1!25}5.16 (86) & \cellcolor{DeepSkyBlue1!25}4.47 & \cellcolor{DeepSkyBlue1!25}5.06 & \cellcolor{DeepSkyBlue1!25}0.32\\ 
\hspace{0.0mm} & & $2^1A^{\prime\prime}$ & 5.26 & 6.53 & 5.54 & 5.01 & 5.99 & 5.52 (83) & 5.83 & 5.80 & 0.21\\ 
\hspace{0.0mm} & & $3^1A'$ & 5.62 & 5.36 & 5.54 & 5.71 & 5.95 & 5.61 (85) & 4.78 & 5.56 & 0.37\\ &&&&&&&&&&& \\[-0.15cm]
\hspace{0.0mm} thymine & 8 & $1^1A^{\prime\prime}$ & 4.82 & 4.39 & 4.94 & 4.94 & 5.14 & 4.98 (87) & 4.91 & 4.90 & 0.23\\ 
\hspace{0.0mm} & & $2^1A'$ & 5.20 & 4.88 & 5.06 & 5.39 & 5.60 & 5.34 (89) & 4.84 & 5.29 & 0.29\\ 
\hspace{0.0mm} & & $3^1A'$ & 6.27 & 5.88 & 6.15 & 6.46 & 6.78 & 6.34 (83) & 4.56 & 5.94 & 0.50\\ 
\hspace{0.0mm} & & $2^1A^{\prime\prime}$ & 6.16 & 5.91 & 6.38 & 6.33 & 6.57 & 6.45 (89) & 6.55 & 6.44 & 0.19\\ 
\hspace{0.0mm} & & $4^1A'$ & 6.53 & 6.10 & 6.52 & 6.80 & 7.05 & 6.71 (88) & 4.91 & 6.45 & 0.48\\ &&&&&&&&&&& \\[-0.15cm]
\hspace{0.0mm} uracil & 8 & $1^1A^{\prime\prime}$ & 4.80 & 4.54 & 4.90 & 4.91 & 5.11 & 4.90 (86) & 4.82 & 4.83 & 0.23\\ 
\hspace{0.0mm} & & $2^1A'$ & 5.35 & 5.00 & 5.23 & 5.52 & 5.70 & 5.44 (88) & 4.80 & 5.36 & 0.32\\ 
\hspace{0.0mm} & & \cellcolor{lightgray}$3^1A'$ & \cellcolor{lightgray}6.26 & \cellcolor{lightgray}5.82 & \cellcolor{lightgray}6.15 & \cellcolor{lightgray}6.43 & \cellcolor{lightgray}6.76 & \cellcolor{lightgray}6.29 (83) & \cellcolor{lightgray}N/A & \cellcolor{lightgray}N/A & \cellcolor{lightgray}N/A \\ 
\hspace{0.0mm} & & $3^1A^{\prime\prime}$ & 6.56 & 6.37 & 6.97 & 6.26 & 6.50 & 6.77 (91) & 5.63 & 6.93 & 0.42\\ 
\hspace{0.0mm} & & $2^1A^{\prime\prime}$ & 6.10 & 6.00 & 6.27 & 6.73 & 7.68 & 6.32 (88) & 6.49 & 6.39 & 0.19\\ 
\hspace{0.0mm} & & $4^1A'$ & 6.70 & 6.46 & 6.75 & 6.96 & 7.19 & 6.87 (88) & 5.14 & 6.40 & 0.43\\ &&&&&&&&&&& \\[-0.15cm]
\hspace{0.0mm} adenine & 10 & $2^1A'$ & 5.25 & 5.13 & 5.20 & 5.28 & 5.37 & 5.18 (86) & 4.41 & 5.36 & 0.41\\ 
\hspace{0.0mm} & & $3^1A'$ & 5.25 & 5.20 & 5.30 & 5.42 & 5.61 & 5.39 (89) & 4.90 & 5.37 & 0.32\\ 
\hspace{0.0mm} & & $1^1A^{\prime\prime}$ & 5.12 & 6.15 & 5.21 & 5.27 & 5.58 & 5.34 (88) & 4.87 & 5.44 & 0.32\\ 
\hspace{0.0mm} & & $2^1A^{\prime\prime}$ & 5.75 & 6.86 & 5.97 & 5.91 & 6.19 & 5.96 (88) & 5.12 & 5.84 & 0.34\\ 
\hline\hline\\
\end{longtable*}

\begin{table*}[htbp]
\caption{Mean unsigned errors and standard deviations for
singlet excitation energies in eV.
States without ESMF solutions and states identified by ESMP2 to have
large doubly excited components
(gray and red rows in Table \ref{tab:all_states_with_tbe})
are excluded.
\label{tab:wfn_red_gray}
}
\resizebox{\textwidth}{!}{
\begin{tabular}{|l|c c c c c c c|}
\hline
\hspace{2mm} & 
\hspace{2mm} SA-CASPT2 & 
\hspace{2mm} MS-CASPT2 & 
\hspace{2mm} CC2 & 
\hspace{2mm} EOM-CCSD & 
\hspace{2mm} CC3 & 
\hspace{2mm} ESMP2 & 
\hspace{2mm} $\epsilon$-ESMP2 \\ \hline
\hspace{0.0mm} Ketones and amides & 0.20 $\pm$ 0.18 & 0.02 $\pm$ 0.05 & 0.29 $\pm$ 0.26 & 0.45 $\pm$ 0.38 & 0.26 $\pm$ 0.31 & 0.39 $\pm$ 0.37 & 0.17 $\pm$ 0.16 \\ 
\hspace{0.0mm} Conjugated polyenes & 0.14 $\pm$ 0.09 & 0.34 $\pm$ 0.34 & 0.32 $\pm$ 0.22 & 0.57 $\pm$ 0.13 & 0.43 $\pm$ 0.12 & 0.27 $\pm$ 0.16 & 0.13 $\pm$ 0.13 \\ 
\hspace{0.0mm} Conjugated rings & 0.32 $\pm$ 0.17 & 0.01 $\pm$ 0.03 & 0.22 $\pm$ 0.07 & 0.36 $\pm$ 0.16 & 0.18 $\pm$ 0.12 & 0.60 $\pm$ 0.40 & 0.15 $\pm$ 0.10 \\ 
\hspace{0.0mm} Heterocycles & 0.34 $\pm$ 0.21 & 0.10 $\pm$ 0.13 & 0.28 $\pm$ 0.19 & 0.43 $\pm$ 0.19 & 0.23 $\pm$ 0.16 & 0.68 $\pm$ 0.42 & 0.17 $\pm$ 0.14 \\ 
\hspace{0.0mm} Nucleobases & 0.41 $\pm$ 0.37 & 0.15 $\pm$ 0.10 & 0.21 $\pm$ 0.13 & 0.47 $\pm$ 0.33 & 0.17 $\pm$ 0.08 & 0.68 $\pm$ 0.52 & 0.19 $\pm$ 0.15 \\ \hline
\hspace{0.0mm} All & 0.31 $\pm$ 0.24 & 0.09 $\pm$ 0.13 & 0.26 $\pm$ 0.18 & 0.43 $\pm$ 0.26 & 0.23 $\pm$ 0.19 & 0.60 $\pm$ 0.43 & 0.17 $\pm$ 0.14 \\ \hline
\end{tabular}}
\end{table*}

\begin{table*}[htbp]
\caption{Mean unsigned errors and standard deviations for
singlet excitation energies in eV.
States without ESMF solutions and states identified by ESMP2 to have
large doubly excited components
(gray and red rows in Table \ref{tab:all_states_with_tbe})
are excluded.
\label{tab:td_dft_red_gray}
}
\resizebox{\textwidth}{!}{
\begin{tabular}{|l|c c c c c c|}
\hline
\hspace{6mm} & 
\hspace{6mm} BP86 & 
\hspace{6mm} B3LYP & 
\hspace{6mm} BHLYP & 
\hspace{6mm} DFT/MRCI & 
\hspace{6mm} ESMP2 & 
\hspace{6mm} $\epsilon$-ESMP2 \\ 
\hline
\hspace{0.0mm} Ketones and amides & 0.55 $\pm$ 0.35 & 0.29 $\pm$ 0.19 & 0.35 $\pm$ 0.44 & 0.34 $\pm$ 0.21 & 0.39 $\pm$ 0.37 & 0.17 $\pm$ 0.16 \\ 
\hspace{0.0mm} Conjugated polyenes & 0.52 $\pm$ 0.32 & 0.40 $\pm$ 0.22 & 0.22 $\pm$ 0.11 & 0.22 $\pm$ 0.13 & 0.27 $\pm$ 0.16 & 0.13 $\pm$ 0.13 \\ 
\hspace{0.0mm} Conjugated rings & 0.51 $\pm$ 0.34 & 0.36 $\pm$ 0.19 & 0.29 $\pm$ 0.22 & 0.16 $\pm$ 0.13 & 0.60 $\pm$ 0.40 & 0.15 $\pm$ 0.10 \\ 
\hspace{0.0mm} Heterocycles & 0.44 $\pm$ 0.29 & 0.21 $\pm$ 0.18 & 0.49 $\pm$ 0.26 & 0.17 $\pm$ 0.12 & 0.68 $\pm$ 0.42 & 0.17 $\pm$ 0.14 \\ 
\hspace{0.0mm} Nucleobases & 0.83 $\pm$ 0.30 & 0.50 $\pm$ 1.20 & 0.57 $\pm$ 0.29 & 0.15 $\pm$ 0.12 & 0.68 $\pm$  0.52 & 0.19 $\pm$ 0.15 \\ \hline
\hspace{0.0mm} All & 0.54 $\pm$ 0.34 & 0.31 $\pm$ 0.54 & 0.44 $\pm$ 0.31 & 0.20 $\pm$ 0.16 & 0.60 $\pm$ 0.43 & 0.17 $\pm$ 0.14 \\ \hline
\end{tabular}}
\end{table*}

\begin{table*}[htbp]
\caption{Mean unsigned errors and standard deviations for
singlet excitation energies in eV.
States without ESMF solutions and states identified by ESMP2 to have
large doubly excited components
(gray and red rows in Table \ref{tab:all_states_with_tbe})
are excluded.
%Mean unsigned errors in eV for molecular groupings based on pi system size. The states that ESMP2 identified as having large doubly excited character and those ESMF did not converge to have been omitted from these values, see the red- and gray-highlighted states in Table \ref{tab:all_states_with_tbe}.
\label{tab:wfnc_pi_red_gray}
}
\resizebox{\textwidth}{!}{
\begin{tabular}{|l|c c c c c c c|}
\hline
\multicolumn{1}{|l|}{$\pi$ system size} & 
\hspace{2mm} SA-CASPT2 & 
\hspace{2mm} MS-CASPT2 & 
\hspace{2mm} CC2 & 
\hspace{2mm} EOM-CCSD & 
\hspace{2mm} CC3 & 
\hspace{2mm} ESMP2 & 
\hspace{2mm} $\epsilon$-ESMP2 \\ \hline
\hspace{0.0mm} 2 & 0.19 $\pm$ 0.17 & 0.11 $\pm$ 0.24 & 0.32 $\pm$ 0.27 & 0.39 $\pm$ 0.36 & 0.28 $\pm$ 0.34 & 0.21 $\pm$ 0.14 & 0.17 $\pm$ 0.16 \\ 
\hspace{0.0mm} 3 & 0.12 $\pm$ 0.11 & 0.00 $\pm$ 0.00 & 0.28 $\pm$ 0.26 & 0.40 $\pm$ 0.43 & 0.30 $\pm$ 0.32 & 0.16 $\pm$ 0.18 & 0.13 $\pm$ 0.09 \\ 
\hspace{0.0mm} 4 & 0.17 $\pm$ 0.16 & 0.17 $\pm$ 0.18 & 0.23 $\pm$ 0.12 & 0.43 $\pm$ 0.16 & 0.29 $\pm$ 0.16 & 0.24 $\pm$ 0.12 & 0.09 $\pm$ 0.13 \\ 
\hspace{0.0mm} 5 & 0.41 $\pm$ 0.13 & 0.10 $\pm$ 0.10 & 0.38 $\pm$ 0.20 & 0.43 $\pm$ 0.18 & 0.21 $\pm$ 0.15 & 0.35 $\pm$ 0.34 & 0.17 $\pm$ 0.14 \\ 
\hspace{0.0mm} 6 & 0.31 $\pm$ 0.22 & 0.11 $\pm$ 0.13 & 0.25 $\pm$ 0.17 & 0.42 $\pm$ 0.20 & 0.24 $\pm$ 0.17 & 0.75 $\pm$ 0.39 & 0.17 $\pm$ 0.14 \\ 
\hspace{0.0mm} 8 & 0.34 $\pm$ 0.26 & 0.11 $\pm$ 0.11 & 0.22 $\pm$ 0.14 & 0.53 $\pm$ 0.31 & 0.19 $\pm$ 0.10 & 0.74 $\pm$ 0.47 & 0.20 $\pm$ 0.16 \\ 
\hspace{0.0mm} 10 & 0.48 $\pm$ 0.34 & 0.04 $\pm$ 0.07 & 0.20 $\pm$ 0.08 & 0.39 $\pm$ 0.13 & 0.17 $\pm$ 0.10 & 0.69 $\pm$ 0.41 & 0.16 $\pm$ 0.10 \\ \hline
\multicolumn{1}{|l|}{5 or less} & 0.24 $\pm$ 0.18 & 0.09 $\pm$ 0.17 & 0.32 $\pm$ 0.23 & 0.41 $\pm$ 0.31 & 0.26 $\pm$ 0.27 & 0.24 $\pm$ 0.23 & 0.16 $\pm$ 0.13 \\ 
\multicolumn{1}{|l|}{6 or more} & 0.34 $\pm$ 0.26 & 0.10 $\pm$ 0.10 & 0.24 $\pm$ 0.15 & 0.45 $\pm$ 0.23 & 0.21 $\pm$ 0.14 & 0.74 $\pm$ 0.41 & 0.18 $\pm$ 0.14 \\ \hline
\end{tabular}}
\end{table*}

\subsection{Comparison to TD-DFT}

Shortly after the introduction of the Thiel benchmark set,
a followup study evaluated the performance of TD-DFT and DFT/MRCI 
on the same molecules and states. \cite{Thiel-BM-TDDFT}
In Table \ref{tab:td_dft_red_gray}, we compare the results of that
study to ESMP2 and $\epsilon$-ESMP2.
Due to its sensitivity to $\pi$ system size, ESMP2 without
regularization is clearly less accurate than typical TD-DFT
approaches, which, having a very different mathematical structure,
do not suffer the same issue of small denominators as the
lowest doubly excited configurations come down in energy.
Indeed, TD-DFT under the usual adiabatic approximation leads to
a formalism in which doubles do not participate in excited states at all.
\cite{HeadGordon:2005:tddft_cis}
$\epsilon$-ESMP2, on the other hand, proves to be more accurate
on the Thiel set singlet states than any of the
TD-DFT functionals originally tested by Thiel,
and this favorable comparison holds even when
considering more recent benchmarking \cite{laurent2013tddft}
of a much wider range of functionals, where MUEs were seen
to range from just above 0.2 eV up to more than 0.5 eV.
Even when the states with large $|\mathrm{T}_2|$ are
included (see tables in SI), $\epsilon$-ESMP2
shows a MUE of 0.19 eV, although it is far from obvious that
such states should be used in comparing these methods
as TD-DFT cannot treat their doubly excited parts at all.
Table \ref{tab:td_dft_red_gray} also shows that
$\epsilon$-ESMP2's accuracy is largely consistent across
different types of molecules, whereas the density functionals
tested by Thiel have accuracies that vary more widely,
with the nucleobases proving the most difficult.

Another difference between TD-DFT and ESMP2 is the latter's
ability to offer diagnostic information about the presence of
doubly excited character.
Although TD-DFT at $N^4$ is less expensive than ESMP2, it offers no
information on such character, whereas ESMP2 can do so at $N^5$ cost.
This is substantially lower than the $N^7$ cost of CC3, and the
original Thiel set study makes clear that lower-level CC methods
like EOM-CCSD are much less effective at predicting doubly
excited character. \cite{Thiel-BM-MAIN}
Thus, when checking for doubly excited character when trying to
assess the trustworthiness of TD-DFT for a particular excited state,
ESMP2 may offer a relatively affordable approach.

\subsection{Group 1: Aldehydes, Ketones, and Amides}

These molecules have many uses as functional groups in biological
and photocatalytic settings,\cite{ba_amide_application,bume_ketone_application,deng_amide_application,gnaim_ketone_application}
making them interesting both from
a formal and a practical perspective. 
Thiel's CASPT2 approach (CASPT2 b) is especially
accurate in this set of molecules with a MUE of just 0.02 eV.
$\epsilon$-ESMP2 is the next most accurate, followed by
Roos's CASPT2, CC3, and CC2, with EOM-CCSD and ESMP2 being
the least accurate.
%, with CC2 and CC3 methods also
%do well with MUEs between 0.2 and 0.35 eV, except, once again, for the $2^1$A$_1$ states, with CC2 and CC3 having slightly smaller errors than EOM-CCSD. In comparison, ESMP2 had a MUE of 0.39 eV and a standard deviation of 0.37 eV, making it most comparable to EOM-CCSD which had an average unsigned error of 0.45 eV and a standard deviation of 0.38 eV.
%When the 0.5 Ha level shift was added to the ESMP2 method ($\epsilon$-ESMP2) the MUE dropped to 0.17 eV with a standard deviation of 0.16 eV, giving it an even lower MUE and spread of errors than one of the two CASPT2 methods listed in the Thiel benchmark. 
% Not sure if this is needed but the perturbative methods tend to error low (though the CASPT2(b) errors are mostly zero for this group) while the coupled-cluster methods tend to error high)
Previous TD-DFT work
shows that TD-DFT methods with hybrid functionals
usually give results comparable to EOM-CCSD
in these molecules.\cite{VanVoorhis2013_ROKS,Trucks2002,Thiel-BM-TDDFT}
%Since Hartree Fock-based methods generally do well for predicting the excitation energies of the excited states for these molecules, functionals with a higher exact exchange (EXX) tend to give more accurate results. While a direct comparison cannot be made with all these benchmarks due to differences in geometry and basis set, it is likely that ESMP2 without a level shift would do similarly well or slightly worse than a well chosen TD-DFT calculation for these molecules.
$\epsilon$-ESMP2 proves to be more accurate than B3LYP in these
molecules, which in turn is interestingly significantly more
accurate than DFT/MRCI, which has more difficulty with this
set of molecules than with any other.
%The Thiel TD-DFT study which used the same geometries and basis set as the calculations presented here, showed that ESMP2 performs similarly to BHHLYP, though BHHLYP tends to error high while ESMP2 errors low. ESMP2 is more accurate than the BP86 functional on average, but less accurate than B3LYP and DTF/MRCI. However, $\epsilon$-ESMP2 performs better than all four methods studied in the benchmark. 

\subsubsection{Formaldehyde and Acetone}

For formaldehyde and acetone three states were studied: $1^1$A$_2$ of $n\rightarrow\pi^*$ character, $1^1$B$_1$ of $\sigma\rightarrow\pi^*$ character, and $2^1$A$_1$ of $\pi\rightarrow\pi^*$ character.
It should be noted that the $2^1$A$_1$ states of these molecules are known to have considerable Rydberg character, which cannot be described properly in the TZVP basis set as it lacks diffuse functions.
We have chosen not to exclude these states from our analysis,
because many other states in the Thiel set also have some
Rydberg character to varying degrees, making it difficult to
draw a clear line between what to include and what not to. 
Of course, all the methods we are comparing with each other
use the same TZVP basis and so are faced with this same issue.
%a basis set without diffuse functions, such as the one used for this study, cannot accurately describe. We will still use these states in our comparisons as all results used in our comparison used the TZVP basis set. %The influence of the basis set on these states is clear as the TBE is based on CC3/aug-cc-pVQZ calculations that are notably incredibly similar to the CC3/TZVP states except for the $2^1$A$_1$ states of these two molecules. 

These two molecules are particularly interesting for ESMP2,
as they are the only cases in this benchmark where ESMP2 did
as well as $\epsilon$-ESMP2.
Both methods produced errors with a relatively small
magnitude of 0.1 eV for the non-Rydberg states.
Interestingly, the addition of the level shift actually
increased errors for the $1^1$B$_1$ state in formaldehyde and
the $1^1$A$_2$ state in acetone, although $\epsilon$-ESMP2
remains quite accurate.
%bringing their errors from the order of 0.01 eV to 0.1 eV. The improvement in error with the addition of the level shift seen in the other two states was minimal compared to this change. However, the errors remain small enough that only the CASPT2 and CC3 methods performed better. EOM-CCSD even incorrectly assigns the order of the states, an issue that our method avoids. 
Another interesting and potentially noteworthy observation we made
was that ESMP2 showed larger doubles norms for the Rydberg states
and a much larger maximum individual amplitude value,
raising the question of whether it would have any value in
flagging Rydberg character.
We don't have enough data in this study to say anything
conclusive on this front, but it may be interesting to
study further.
%While the Rydberg states will be discussed here even though the TZVP basis set lacks the diffuse functions needed to properly describe these states, it is important to note that our method did flag these two states as results that should not be trusted. The singles percentages calculated from the first order wave function for the $2^1$A$_1$ states were 3\% lower than the norms of the other states studied for the same molecule, showing that the a larger perturbation was needed to describe these states. Our method also did not fully replicate the singles vector generated by EOM-CCSD for the $2^1$A$_1$ state of acetone, casting further doubt that ESMP2 method provides a good description for the state. 

\subsubsection{\textit{p}-Benzoquinone}

%fun fact! about scorpions! and cancer! We don't have to include this just wanted to jazz it up.
%Para-benzoquinone is interesting for a variety of reasons but recently has been investigated as a motif for biologically active molecules with antimicrobial\cite{Zare2019_benzoquinone-scorpions} and anticancer properties\cite{Srinivas2004_benzoquinone-cancer} and, more relevant to its excited state properties, for bioimaging purposes.\cite{Dias2018_benzoquinone-bioimaging}

For benzoquinone, three $n\rightarrow\pi^*$ states - $1^1$A$_u$, $1^1$B$_{1g}$, $1^1$B$_{3u}$ - and three $\pi\rightarrow\pi^*$ states - $1^1$B$_{3g}$, $1^1$B$_{1u}$, $2^1$B$_{3g}$ - were studied. 
Within this group of molecules, benzoquinone showed by far the
largest amount of doubly excited character, as seen in both the
CC3 \%T\textsubscript{1} and the ESMP2 $|\mathrm{T}_2|$ values.
Unsurprisingly, unregularized ESMP2 performed quite poorly
in benzoquinone, with $\epsilon$-ESMP2 performing much better
and more comparably to CC2 and CC3.
%was the only one to have enough doubly excited character to impact the ESMP2 calculations. Based on the  values $1^1$B$_{3u}$ had below 80\% singles character, $1^1$A$_u$ and $1^1$B$_{1g}$ were slightly higher at around 84\% and the remaining states had around 89\% singles character. Note that none of these percentages were above 90\%. As our method only accounts for single excitations in the reference method, any state with non-negligible amounts of doubly excited character will cause the perturbation needed to describe the state to not be ``small." This is clearly seen when observing the ESMP2 errors compared to the TBEs taken from CASPT2 calculations: all errors were above 0.5 eV with the lowest being 0.58 and 0.59 eV for the $1^1$B$_{3g}$ and $1^1$B$_{1u}$ states and the largest errors were 1.03 and 1.09 eV for the $1^1$A$_u$ and $1^1$B$_{1g}$ states. 
%While the singles norms calculated from the first order ESMP2 wave function do generally decrease as the errors of ESMP2 increase, it is important to note that the lowest singles norm was for the $1^1$B$_{3u}$ state which had an error in the middle of the range for this molecule with a magnitude of 0.72 eV. Based on ESMP singles percentages from similarly sized molecules (octatetraene, cytosine) the percentages for the benzoquinone states were all relatively low, as a state dominated by single excitations can correspond to percentages as high as 82\%. 
%Adding the level shift to ESMP2 reduced all but one of the errors to below 0.3 eV. The one remaining
$\epsilon$-ESMP2's largest error in this molecule 
was 0.54 eV for the $1^1$B$_{3u}$ state, which has the most
significant doubly excited character.
This reminds us that, although $\epsilon$-ESMP2 can improve
significantly over ESMP2 when such character is present,
it is no substitute for multi-reference methods in cases
where the doubly excited component is large enough.
%the state with the largest doubly excited character by far, thus it is unsurprising that even with an aggressive level shift our method could not produce a highly accurate result. However, this was not the only state with a small ESMP singles percentage after the level shift was added, $2^1$B$_{3g}$ had a singles percentage that was only slightly bigger at 84.2\% compared to 84.0\%, and yet $\epsilon$-ESMP2 had an error of only 0.03 eV. This state, based on CC3 data, also did not have an unusually large amount of doubly excited character compared to the other states here, so it is unclear why $\epsilon$-ESMP2 still produces a relatively small singles percentage. 
%Compared to the other wave function-based methods, ESMP2 performs worse than the other methods with results most similar to those from EOM-CCSD. The sharp reduction in error brought by adding the level shift makes $\epsilon$-ESMP2 more comparable to CC2, CC3, and CASPT2. 

\subsubsection{Formamide, Acetamide, and Propanamide}

For each of these molecules, the $1^1$A$^{\prime\prime}$
$n\rightarrow\pi^*$ state, and the
$2^1$A$^\prime$ $\pi\rightarrow\pi^*$ state were studied.
In the $2^1$A$^\prime$ state of propanamide
and especially acetamide, the excitation within the converged
ESMF wave function contained fewer components than in EOM-CCSD,
placing a higher fraction of the overall weight on the
dominant HOMO$\rightarrow$LUMO+2 component.
We see this as a good reminder that both orbital relaxation
and the degree to which correlation effects are captured can
affect the degree of predicted mixing between excitation components.
%As with the other states studied, we used the singles vector from EOM-CCSD to help determine if our method had converged to the targeted state. While ESMF was able to converge to a state with a CI vector that closely matched the vectors from EOM-CCSD for the $1^1$A$^{\prime\prime}$ states for these three molecules, the vectors from the $2^1$A$^\prime$ states did not match as well. For these states EOM-CCSD predicted that there would be a large contribution from one single excitation and then a slightly smaller contribution from another single excitation for these states. ESMF was able to converge to stationary points that captured the character only from the larger of the two contributions. 

ESMP2 was quite accurate for the $1^1$A$^{\prime\prime}$ excitation
energies, with $\epsilon$-ESMP2 less so, while
$\epsilon$-ESMP2 was much more accurate than ESMP2 for the
$2^1$A$^\prime$ states.
%, with errors less than 0.05 eV for formamide and propanamide and an error of 0.14 eV for acetamide. ESMP2 had an error of only 0.02 eV for the $2^1$A$^\prime$ state of formamide, however, the $2^1$A$^\prime$ states of acetamide and propanamide were much less accurate with errors of around 0.4 eV.
Although none of these states has a particularly high degree
of doubly excited character, the ESMP2 doubles norms do correctly
predict the relative accuracy for the unregularized theory
between these two states.
%Comparing the singles norms between the two states presented for each molecule shows that ESMP2 predicts that the results should be less accurate for the $2^1$A$^\prime$ states for all three molecules as there is a lowering of around 3\% in each case. The small error in formamide's $2^1$A$^\prime$ state is likely coming from coincidental cancellation of errors. Looking at the CC3 data, a possible reason for the lack of ESMP2 to accurately calculate the excitation energies of these states could be that these states have slight doubly excited character as the T\textsubscript{1}\% values are all below 90\%. The other wave function based methods, except for CASPT2, showed similar trends in errors between the states of A$^\prime$ and A$^{\prime\prime}$. 
A final noteworthy point is the unusually high errors made by
CC2, EOM-CCSD, and CC3 in the $2^1$A$^\prime$ state of
formamide.
It is not obvious what is driving this error, especially
considering ESMP2's accuracy and the small effect of
introducing regularization.

%Adding the 0.5 level shift to the ESMP2 method unfortunately increased the errors for the $1^1$A$^{\prime\prime}$ states by around 0.15 eV in all three cases. However, the errors for the $2^1$A$^\prime$ states in acetamide and propanamide were decreased by over 0.3 eV, so on average adding the level shift increased the accuracy of these results. 

\subsection{Group 2: Conjugated Polyenes}

The four unsaturated polyene molecules in this group
-- ethene, butadiene, hexatriene, and octatetraene --
provided a great deal of insight into how ESMP2 performs in the
presence of doubly excited character, as the $2^1$A$_g$ states of
butadiene, hexatriene, and octatetraene all have large
doubly excited components, \cite{nakayama1998polyenes}
which can for example be seen in their CC3 \%T$_1$ values.
The ESMP2 doubles norm correctly flags all three of these
doubly excited $2^1$A$_g$ states as likely to be problematic
for ESMP2 and other single-reference methods.
%For all averages/standard deviations discussed these doubly excited states are excluded as ESMF/ESMP2 does not have the ability to describe doubly exited character due to the wave function ansatz.
Although far superior to the other methods in the doubly
excited state, Thiel's CASPT2 results (CASPT2 b) are not
especially competitive for the 1$^1$B$_2$ states.
Even more surprising is the degree of difficulty
that CC3 has with the 1$^1$B$_2$ states, as they are
dominated by singly excited components.
%When looking at the mean unsigned errors and standard deviations for this group of molecules, ESMP2, with an MUE of 0.27 eV and a standard deviation of 0.16 eV performs most similarly to CC2 which had an MUE of 0.32 eV and a standard deviation of 0.22 eV. ESMP2 does significantly better than EOM-CCSD and marginally better than CC3 for these excited states. The ESMP2 MUE and standard deviation were between the two CASPT2 values presented in the 2008 Thiel benchmark, with the calculations with the smaller (Roos) active space performing better than the more sophisticated MS-CASSCF method. This is likely from a convenient cancellation of errors, as the larger active space used in the MS-CASSCF method should provide a more complete picture for these states. Adding the 0.5 Ha level shift reduced the MUE and standard deviation of the ESMP2 method to 0.13 eV, making it one of the best wave function-based methods out of those discussed here. 
%Looking at TD-DFT results for these molecules, it appears that basis set choice can greatly determine the accuracy of the results predicted for these methods. %ROKS paper
Excited-state-specific DFT in the form of restricted
open-shell Kohn Sham (ROKS) has also shown difficulty
in these states, \cite{VanVoorhis2013_ROKS} with
accuracy appearing to decrease as the basis set is enlarged.
In the TD-DFT benchmark presented by Wiberg et al.,
\cite{Trucks2002}
it was shown that the ethene and butadiene singly excited
states were modeled best by functionals with higher amounts
of Hartree Fock exchange.
Indeed, in Thiel's TD-DFT benchmark, \cite{Thiel-BM-TDDFT}
BHLYP significantly outperformed B3LYP and BP86 in the
1$^1$B$_2$ states, although we find that
$\epsilon$-ESMP2 does better still.
%did similarly to ESMP2 without a level shift, though B3LYP and DFT/MRCI had a smaller spread of errors. BP86 had a similar spread of errors but a much higher MUE than ESMP2. The decrease in error produced by adding the level shift to ESMP2 made $\epsilon$-ESMP2 produce results significantly better than the DFT results presented in this benchmark. 

\subsubsection{Ethene}

The only state studied for ethene was the $1^1$B$_{1u}$ state as the other low lying excited states for the molecule are strongly Rydberg in character and cannot be accurately described using the TZVP basis set used here.\cite{krauss_mielczarek_ethene,roos_ethene} The $1^1$B$_{1u}$ state also contains significant valence-Rydberg mixing, however, it is still mostly described as a valence excited state. The Thiel best estimate value of 7.80 eV is based on a mixture experimental data and high-level \textit{ab initio} results, though it was noted in the paper that defined the best estimates that the vertical excitation of the $1^1$B$_{1u}$ state could not be assigned precisely based on experimental data.\cite{Thiel-BM-MAIN}

In this state, ESMP2 and $\epsilon$-ESMP2 performed similarly with errors of 0.25 and 0.24 eV, respectively, which makes sense given
how strongly dominated this state is by single excitations.
%While there are no other states for ethene that we can compare the singles amplitude norm against, we can compare this percentages to those of formaldehyde as the two molecules have the same number of electrons. Based on this comparison the singles norm of 94\% for the $1^1$B$_{1u}$ state suggests that ESMP2 performs reasonably well.
The CC methods EOM-CCSD, CC2, and CC3 all perform relatively poorly
for this state with errors around 0.6-0.7 eV,
and Thiel's CASPT2 shows an unusually large error of 0.82 eV.
Roos's CASPT2 error is much smaller at 0.18 eV,
and Thiel et al.\ report an almost exact result with a greatly
expanded (8,20) active space, \cite{Thiel-BM-MAIN}
a useful reminder of how
important the choice of active space can be.

\subsubsection{Butadiene, Hexatriene, and Octatetraene}

%{\color{red} Cite Trucks's work on butadiene.\cite{Trucks2002} this looks at the excited states of ethene, isobutene, formaldehyde, and acetone calculated using RPA, TDDFT, EOM-CCSD. Not sure if any information from this study actually needs to be included since it's using a pople basis set, making comparisons to the values here difficult.}

For these three molecules, two states were studied each:
the single-excitation-dominated $1^1$B$_u$ state,
and the substantially doubly excited $2^1$A$_g$ state.
As expected, ESMP2 performs considerably better for the $1^1$B$_u$ states
(errors of 0.5 eV or below) than for the $2^1$A$_g$
(errors between 2 and 3 eV).
Its redeeming quality in the latter states is it's ability to signal
its own failure through unusually large doubles norms of 0.77, 1.0, and
1.15, clearly warning the user to get their hands on a multi-reference
method instead.
To put how extreme these norms are in context, remember that the weight
of the zeroth order reference in intermediate normalization is 1,
meaning that this perturbation theory's perturbation is coming out
as big or bigger than the zeroth order piece!
Such a grossly nonsensical result is a clear sign of failure, which
if heeded can help guide a user in selecting a more appropriate method.
%The singles norms from ESMP2 do correctly show that ESMP2 should not be used for these states A$_g$ symmetry states: their values were 60\%, 47\%, and 40\% for butadiene, hexatriene, and octatetraene, respectively. These are well below the 80-90\% that would be expected for a largely singly excited state. These are errors versus the Thiel best estimates, which came from MR-CI based calculations in butadiene and multi-reference M\o ller-Plesset perturbation theory (MRMP) for hexatriene and octatetraene. It should be noted that the ordering of the two singlet states in butadiene is somewhat unclear and that the theoretical best estimates for the energy of the $2^1$A$_g$ states for all three molecules has not been verified experimentally.
%{\color{blue} Not sure if this needs to be said, but the poor ability of ESMF to replicate the wave function for the 21Ag states resulted in all three states of that symmetry label to have non-zero overlap with the ground state.}
As for the $1^1$B$_u$ state, because our summary tables exclude states
flagged as strongly doubly excited by ESMP2,
the entry in Table \ref{tab:wfn_red_gray} offers at a glance the
performance on this less challenging, singly-excited state.
$\epsilon$-ESMP2 is considerably more accurate for this state than
the CC methods, rivaled only by CASPT2 approaches with well chosen
active spaces.
%for comparison against other wave function-based methods, for butadiene and hexatriene ESMP2 without a level shift performs incredibly well, with errors lower than all other methods shown except for SA-CASPT2. ESMP2 predicted an error 0.5 eV for the $1^1$B$_u$ state of octatetraene, an error larger in magnitude than the other methods studied here. The magnitude of the error from the octatetraene result unique compared to the same states of butadiene and hexatriene, however, this relatively large error was seen in the results of the other wave function-based methods as well. In CC2, CC3, and MS-CASPT2 the octatetraene error was comparatively low while in EOM-CCSD and SA-CASPT2 it was oddly high. This could be because the inclusion of doubly (and higher) excited character is slightly more important for this state or it is possible that the MRMP value taken as the best estimate is not as good of an estimate for this state as the TBEs were for the butadiene and hexatriene states. 

\subsection{Group 3: Conjugated rings}

%Based on the singles amplitude norms calculated from the first order ESMP2 wave function, ESMP2 should not be used to study the $2^1$A$_1$ state of cyclopentadiene, the $1^1$E$_{2g}$ state of benzene, and the $2^1$B$_{1g}$ state of naphthalene as these states had norms below 70\%. ESMF was unable to converge to a state that resembled the $3^1$A$_g$ state. While these states will still be discussed individually in the following sections, they will be excluded from any averages taken to compare the methods discussed here. 

With their larger $\pi$ systems, this group of molecules proved
especially difficult for unregularized ESMP2, whose overall accuracy
in this group was worse than the other wave function methods.
$\epsilon$-ESMP2, on the other hand, outperformed the CC methods,
and was in turn outperformed by Thiel's CASPT2.
Although Thiel's selecton of CASPT2 
to be the TBE in cyclopropene, norbornadiene, and naphthalene no
doubt gives it a statistical advantage, we certainly expect it to be
more accurate than $\epsilon$-ESMP2 in these molecules.
%Based on mean unsigned errors, ESMP2 had the worst accuracy for these molecules with an MUE of 0.54 eV. EOM-CCSD had the next highest MUE of 0.35 eV followed by SA-CASPT2 with 0.30 eV, then CC2 with 0.21 eV, and then CC3 with 0.18 eV. MS-CASPT2 performed the best with an MUE of 0.02, however this result is slightly misleading as the TBEs for cyclopropene, norbornadiene, and naphthalene are directly taken from the MS-CASPT2/TZVP result, resulting in an artificially lowered MUE for this method. Adding a level shift of 0.5 Ha to ESMP2 resulted in an MUE of 0.15 eV, making it the most accurate (on average) of all the methods, except for MS-CASPT2. %EOM, CC2, and CC3 tended to overestimate the energies while CASPT2 and ESMP2 underestimated the energies. Adding the level shift to ESMP2 caused the method to still on average underestimate the energies, but much less so than before and some energies were actually overestimated. -
%Many of the previously cited TD-DFT studies do not include these molecules. %Probably should try to find a few more to cite. 
As in the polyenes, ROKS has shown a tendency for its accuracy to
decrease with increasing basis set size in a number of these conjugated rings,
both with and without the use of range separation.
\cite{VanVoorhis2013_ROKS}
%paper did study the lowest energy excitation for each of these molecules except for naphthalene and showed that for molecules in which the energy was below the best estimate (cyclopropene, cyclopentadiene, norbornadiene) increasing the size of the basis set from 6-31G* to 6-311+G* actually increased the error of the DFT prediction as it further lowered the calculated excitation energy. For the studied benzene excited state, DFT with a 6-31G* basis set overestimated the excitation energy so increasing the basis set lowered the excitation energy, thus decreasing the error.
%Going from a density functional that does not have range separation to one that does causes an increase in energy for all molecules of about 0.1 eV, this has less impact on the predictions than changing the basis set.
DFT/MRCI, on the other hand performs quite well in these molecules,
as does CC3.
%as with an MUE of 0.15 eV, about half the MUEs of BHLYP and B3LYP. BP86 once again on average was the least accurate and had the widest spread of errors based on the standard deviation. ESMP2 without a level shift performed worse than BP86, however the addition of the level shift brought the MUE to 0.15 eV, matching that of DFT/MRCI for this set of molecules and with a slightly smaller standard deviation than DFT/MRCI.

\subsubsection{Cyclopropene}

For cyclopropene two states were studied, the $1^1$B$_1$ $\sigma\rightarrow\pi^{*}$ state and the $1^1$B$_2$ $\pi\rightarrow\pi^{*}$ state. The Thiel best estimate for these states comes directly from a MS-CASPT2/TZVP calculation as it gives a more accurate description for the valence-Rydberg mixing in the $1^1$B$_2$ state even though these values are slightly higher than experiment.\cite{Thiel-BM-MAIN,roos_cyclopropene}
Neither of these states shows any reason for great concern for ESMP2,
and it offers reasonable accuracy that is slightly improved in both
cases by regularization.
Unsurprisingly, CC2 and especially CC3 work well in these states.
%Based on the singles amplitude norms for the two states, ESMP2 should be equally accurate for both states as both had a norm of 91\%. While these norms are high, it should be noted that the two other molecules of similar size in this benchmarking set, formaldehyde and ethene, had norms of around 94\% for states that ESMP2 had extremely accurate predictions for. Based on this it is unsurprising that ESMP2 without a level shift had errors of 0.2 and 0.31 eV for the $1^1$B$_1$ and $1^1$B$_2$ states respectively. As stated, there is valence-Rydberg mixing in the $1^1$B$_2$ state that might be contributing to the slightly higher error. Compared to other wave function based methods, ESMP2 is among the best methods for the $1^1$B$_1$ state but one of the worst for the $1^1$B$_2$ state. Adding the level shift only mildly decreases the errors to 0.15 and 0.21 eV for the $1^1$B$_1$ and $1^1$B$_2$ states, respectively, which did not appreciably change how ESMP2 performed with respect to the other methods. 

\subsubsection{Cyclopentadiene}

We look at two $\pi\rightarrow\pi^*$ excitations in cyclopentadiene -- the $1^1 B_2$ and $2^1 A_1$ states. While both are valence excited states without significant Rydberg mixing,\cite{Roos_CASSCF_cyclopentadienes} the former state is dominated by a single excitation, while the latter is a superposition of components that includes doubly excited pieces.
Thiel calculates a 5.55 eV excitation energy from EOM-CCSDT evaluated with an ``exhaustive''
basis set as the TBE for the $1^1 B_2$ state, which is
a bit above the carefully estimated 5.43(5) eV experimental value.
\cite{Jurgen2004_expcyclopentadiene}
For the $2^1 A_1$ state, Thiel uses CASPT2 for the TBE.
%It is very close to the EOM-CCSD(T) result, which lies at 6.37 eV.\cite{Bartlett1996_eth-but-cyc}
ESMP2 successfully predicts its own failure in the $2^1 A_1$
state, while $\epsilon$-ESMP2 is similar in accuracy to CC3.
As expected, ESMP2 does better for the $1^1 B_2$ state, and
in that case is further improved by $\epsilon$-ESMP2, which
errors low by about the same amount that CC3 errors high.

%Looking at the ESMP2 results, the norms of the first order wave function amplitudes showed that ESMP2 should be much more accurate for the $1^1$B$_2$ state than the $2^1$A$_1$ state with the norm for the $2^1$A$_1$ state being so low (70\%) as to make clear that ESMF cannot describe this state well enough to allow for a perturbative correction. This inability to describe the state as a sum over single excitations caused the converged result to have a nonzero overlap with the ground state. The small amplitude norm for the $2^1$A$_1$ state also agrees with our knowledge of doubly excited character within this state. However, even with the much larger norm of 88\% for the $1^1$B$_2$ state, ESMP2 with out a level shift is still lacking in accuracy when compared to the other wave function-based methods. It ties with EOM-CCSD for the highest unsigned error. For the $2^1$A$_1$ state ESMP2 had an error of 1.48 eV, which is unsurprisingly the worst out of the methods studied here, however it should be noted that CC2 and EOM-CCSD also perform poorly showing that higher excited character is definitely present in the description of this state. Adding the level shift reduced the errors to 0.19 eV and 0.29 eV, for the $1^1$B$_2$ and $2^1$A$_1$ states respectively making $\epsilon$-ESMP2 on par with the most accurate methods for this molecule. 

\subsubsection{Norbornadiene}

Norbornadiene can be seen as the third and most structurally
complicated member of a series begun by
cis-butadiene and cyclopentadiene.
\cite{Roos_CASSCF_norbornadiene} 
For calculations on the excited states of norbornadiene, one must consider
that while it is formally not conjugated, there is indirect conjugation
of the double bonds, allowing for through-space and through-bond
interactions -- thus, interactions between $\pi$ and $\sigma$ orbitals
are more important. \cite{Roos_CASSCF_norbornadiene} 
Two $\pi\rightarrow\pi^{*}$ excitations are examined in this benchmark
-- an experimentally forbidden $1^1$A$_2$ state and a $1^1$B$_2$
excited state that can mix strongly with nearby
Rydberg states. \cite{Roos_CASSCF_norbornadiene}
CASSCF studies, \cite{Roos_CASSCF_norbornadiene} CC3, and ESMP2
all indicate that both states are dominated by single excitations.
Thiel selects CASPT2/TZVP for both TBEs, 5.34 eV and 6.11 eV,
which lie a little above the reported
experimental values of 5.25 eV and 5.95 eV.
\cite{McDiarmid1981_norbornadiene} 

%The first order wave function norms for the two states were above 80\% which seems fairly typical compared to other states described by mostly single excitations for molecules of this size.
ESMP2 produces excitation energy errors of 0.25 eV and 0.32 eV
for the $1^1$A$_2$ and $1^1$B$_2$ states, respectively, making it
more accurate than EOM-CCSD and on par with CC2 and CC3.
$\epsilon$-ESMP2 is the most accurate non-active-space method
for these states, but is not as accurate as CASPT2.
%, though it is worth remembering that the TBEs for this molecule are based on MS-CASPT2/TZVP results. Adding the level shift to ESMP2 drastically reduced the error of the $1^1$A$_2$ state to 0.03 eV and the $1^1$B$_2$ state to 0.20 eV. In terms of accuracy, this makes $\epsilon$-ESMP2 definitively more accurate than CC2 and CC3 for these two states, but still less accurate than CASPT2. 

\subsubsection{Benzene} 

For benzene, we looked at the $1^1$B$_{1u}$, $1^1$B$_{2u}$, $1^1$E$_{1u}$, and $2^1$E$_{2g}$ $\pi\rightarrow\pi^*$ excitations.
The first three of these excitations are dominated by equally-weighted
superpositions of excitations out of the degenerate $\pi$
system HOMOs, while the largest component of the $2^1$E$_{2g}$ state
is a single excitation out of the lowest energy $\pi$ orbital. 
Thiel adopts  S{\'{a}}nchez de Mer{\'{a}}s et al.'s CC3/ANO1
results as best estimates for the states.
\cite{SanchezDeMeras1996_CC_benzene} 
%{\color{red}DO WE NEED TO MENTION THIS? Due to symmetry limitations of the quantum chemistry packages used in this work, our EOM-CCSD calculations used the D${}_{2h}$ symmetry point group, for which the symmetry of the 1 ${}^1 B_{1u}$ state becomes $B_{3u}$, and the symmetry of the 1 ${}^1 B_{2u}$ state becomes $B_{2u}$, while we opted to use the second state of $B_{3u}$ symmetry for the 1 ${}^1 E_{1u}$ excitation and the second state of $A_g$ symmetry for the 2 ${}^1 E_{2g}$ excitation.}
%None of this is about ESMP2, ony about the methods presented in the Thiel benchmark so I'm not sure it really needs to be said here. Leaving it in for now.
%According to their benchmark of select coupled cluster methods,
%\cite{SanchezDeMeras1996_CC_benzene} calculations utilizing the ANO1 basis set show high singly-excited character in the $1^1$B$_{1u}$, $1^1$B$_{2u}$, and $1^1$E$_{1u}$ excitations CCSD wave functions that does not change considerably in CC3 -- however,

Benzene is a good example of a molecule where the EOM-CCSD can
overestimate the degree of singly excited character compared
to CC3.
This issue is particularly stark in the
$2^1$E$_{2g}$ state, where EOM-CCSD and CC3 disagree in
their \% T1 measures by 19\%.
Similarly, the inclusion of triples drops CC3's excitation energy
in this state by 0.6 eV compared to CC2.
%and inclusion of triples in CC3 decreases the excitation energy by 0.765 eV. 
%This result is also reflected by Thiel's results, where EOM-CCSD/TZVP overestimates the excitation energy by 0.800 eV relative to the CC3/TZVP best estimate.\cite{Thiel-BM-MAIN}
%
%Additionally, S{\'{a}}nchez de Mer{\'{a}}s and co. do not consider the CCSD singly excited character of 85\% in the 2 ${}^1 E_{2g}$ state to be so small such as to preemptively suggest that including triples would have such a large effect on the character and energy of the excitation.\cite{SanchezDeMeras1996_CC_benzene}
%
As pointed out by S{\'{a}}nchez de Mer{\'{a}}s et al,
\cite{SanchezDeMeras1996_CC_benzene} 
not all of these states show a uniform convergence order
between CCS, CC2, EOM-CCSD, and CC3, with EOM-CCSD's excitation
energy in the 1$^1B_{2u}$ state lying below that of CC2,
which is atypical among Thiel set states.
%They also point out an interesting deviation in the convergence of the CCS, CC2, CCSD, and CC3 series for benzene -- as these methods sequentially recover higher order descriptions of double excitations in the ground state fluctuation potential, with CCS not describing doubles at all, to CC3 having second-order effects of double excitations, one would expect the excitation energy to uniformly converge as one moves from CCS, to CC2, to CCSD, and then finally to CC3.\cite{Olsen1996_CC_series_converge} However, this trend is not present in the benchmarking data shown. With the exception of the $1^1$B$_{2u}$ state, there is not uniform convergence with respect to the CC3 excitation energy across the CCS, CC2, and CCSD methods in S{\'{a}}nchez de Mer{\'{a}}s and co.'s study,\cite{SanchezDeMeras1996_CC_benzene} or across CC2, CCSD, and CC3 in Thiel and co.'s study.\cite{Thiel-BM-MAIN}
%
These authors go on to use benzene to support an argument that
%From this data, S{\'{a}}nchez de Mer{\'{a}}s and co. caution against trusting large singly excited weights of
EOM-CCSD is not a reliable guide to doubly excited character.
%wave functions as evidence of the unimportance of higher excitations. 
%
With such considerable changes when going from the
inclusion of doubles to the inclusion of triples,
we have some doubts about the accuracy of the CC3 results
as a best estimate for the significantly doubly excited
$2^1$E$_{2g}$ state, and might instead have adopted CASPT2 values.

As in many other cases, ESMP2's doubles norms tell a similar
story about doubly excited character as the CC3 \% T1,
and in particular signal clearly that ESMP2 is not appropriate
for use in the $2^1$E$_{2g}$ state.
ESMP2's excitation energy accuracy is poor in all
of benzene's states, as is common for systems with six or more
orbitals in their $\pi$ system, but, with the exception of
the $2^1$E$_{2g}$ state, $\epsilon$-ESMP2 makes
a large improvement to the point that it is competitive
with CC2 and CC3.
In $2^1$E$_{2g}$, where ESMP2 signals its own failure,
it is difficult not to look to CASPT2 as the preferred
method among those tested, all things considered.

\subsubsection{Naphthalene}

We looked at the $1^1$B$_{3u}$, $1^1$B$_{2u}$, $2^1$A$_g$, $1^1$B$_{1g}$, $2^1$B$_{3u}$, $2^1$B$_{1g}$, $2^1$B$_{2u}$, and $3^1$A$_g$ states of naphthalene, all of which are $\pi\rightarrow\pi^*$ transitions. The TBEs for these states were taken directly from the MS-CASPT2/TZVP results, as
this molecule's size limits other options.
%MS-CASPT2 tended to predict slightly higher excitation energies than the states for which gas phase experimental results are available, all of the states of \textit{gerade} symmetry do not have experimental values. 
Based on the T\textsubscript{1}\% values from CC3, it is likely
that many of these states involve significant amounts of
doubly excited character.
%Exceptions are the $1^1$B$_{2u}$, $2^1$B$_{3u}$, $2^1$B$_{1g}$, and $2^1$B$_{2u}$ states as they had T\textsubscript{1}\% values above 90\&.
The $3^1$A$_g$ state had a particularly low T\textsubscript{1}\% of 70\%,
and is the first state we come to for which the ESMF stationary point
could not be found. 

For the other states, ESMP2 displayed a range of accuracies.
It showed a particularly small error of 0.04 eV for the
$1^1$B$_{1g}$ state, which, despite a relatively low CC3
T\textsubscript{1}\%, was also treated accurately by CC3
and CC2.
In fact, this is one of the rare states in which regularization
made the excitation energy prediction worse, with $\epsilon$-ESMP2
giving an error of 0.37 eV.
%and this state had one of the larger singles norms for the naphthalene states of 76\%. As Naphthalene is comparable in size to adenine, based on the singles norms for these two molecules, it seems as though a norm of 75\% or above will correspond states that are largely dominated by single excitations for molecules of this size.
ESMP2's errors in the other naphthalene states were much larger,
with some states showing significantly larger doubles norms as well,
although not as large as in the doubly excited polyene states.
In these other states, regularization makes a large improvement,
making $\epsilon$-ESMP2 competitive with CC2 and CC3.
%For the other states smaller norms (71-73\%) did correspond to larger errors, with a small exception coming from the $2^1$B$_{3u}$ state which had an error of 0.74 eV but a singles norm of 75\%. $2^1$B$_{1g}$ had a singles norm below 70\% and correspondingly had the largest error out of the naphthalene states of 1.52 eV. Adding the level shift lowered all of the errors for these states to below 0.2 eV with the exception of the $2^1$A$_{g}$ state with an error of 0.21 eV and the $1^1$B$_{1g}$ state with an error of 0.37 eV. The large error for the $1^1$B$_{1g}$ state may be from the nonzero overlap between this state and the $2^1$B$_{1g}$ state. Even the error for the $2^1$B$_{1g}$ state was reduced to only 0.18 eV. ESMP2 without the addition of a level shift performed poorly compared to the other wave function based methods for these states, in terms of accuracy it was most similar to EOM-CCSD and SA-CASPT2. $\epsilon$-ESMP2, however, was the most accurate method (aside from MS-CASPT2 which the TBEs were directly taken from), though CC3 and CC2 had errors only slightly larger in magnitude. 

\subsection{Group 4: Heterocycles}

The molecules in this group are furan, pyrrole, imidazole, pyridine, pyrazine, pyrimidine, pyridazine, triazine, and tetrazine.
A common theme in these molecules is that almost all of the states
studied here have at least moderate contributions from double excitations,
at least as measured by the CC3 T\textsubscript{1}\%.
%and higher excitations which cannot be accurately replicated by many of the methods studied here.
As one might therefore expect, ESMP2's predictions were fairly inaccurate
in this group.
%for these states as the reference, ESMF, does not have double excitations in its wave function ansatz.
Regularization via $\epsilon$-ESMP2 dramatically reduces these errors,
to the point that it is more accurate than CC2, EOM-CCSD, and CC3.
As in many other cases, only CASPT2 with Thiel's active spaces did better.
%with the mean unsigned error going from 0.72 eV to 0.19 eV. This drastic change led us to theorize that the impact of the level shift on ESMP2 was to lower the incorrectly large amount of energy coming from the double and triple excitation terms in the ESMP first order wave function. 
%Compared to other methods, ESMP2 was the least accurate followed by EOM-CCSD with an MUE close to half that of ESMP2. Next lowest in accuracy for this group was SA-CASSCF with an MUE of 0.34 eV, followed by CC2 with 0.28 eV, then CC3 with 0.23 eV, then $\epsilon$-ESMP2, and finally MS-CASPT2 with an MUE of 0.10 eV. 
It should, however, be noted that the CASPT2 b error is somewhat artificially
small in this group, as it was used as the TBE for imidazole, pyridazine,
triazine, and tetrazine. \cite{Thiel-BM-MAIN} 
Among DFT approaches, DFT/MRCI and TD-DFT/B3LYP perform particularly well
in the heterocycles, \cite{Thiel-BM-TDDFT} while
%Looking at existing TD-DFT benchmarks showed that for the lowest excitations for each of these molecules,
ROKS performs well for some cases but shows difficult
basis set dependence in others. \cite{VanVoorhis2013_ROKS}
Further analysis of TD-DFT in some of these
molecules can be found in a study by Caricato, et al. \cite{Wiberg2010}

\subsubsection{Pyrrole and Furan}

%Of particular interest due to their prevalence as motifs in biomolecules\cite{HeterocyclesNaturalProducts_Furan,HeterocyclesNaturalProducts_Pyrrole} and organic polymers,\cite{Gandini2010_furan-polymer,Muller2011_PyrroleConductivePolymer} we now turn to the heterocyclic pentadienes pyrrole and furan.
We consider the following $\pi\rightarrow\pi^{*}$ excitations in pyrrole and furan: the 
1${}^1$B${}_2$, 
2${}^1$A${}_1$, and 
3${}^1$A${}_1$ states.
We analyze these two molecules together as their spectra are similar. 
Pyrrole's  2${}^1$A${}_1$ state and 
furan's 1${}^1$B${}_2$ and 2${}^1$A${}_1$ states are valence excited states, 
while the remaining states are considered to have Rydberg character.
\cite{Roos_CASSCF_cyclopentadienes}
For both molecules, Thiel's TBEs are based on CC3 calculations
with basis set corrections.
\cite{Jorgensen1999_pyrrole,Jorgensen1998_furan}

The ESMP2 predictions were mixed in terms of accuracy. Both molecules had two states, the $3^1$A$_1$ and $1^1$B$_2$ states, that produces errors lower than 0.5 eV and then a $2^1$A$_1$ state with an error of 1.53 eV in furan and 1.07 eV in pyrrole.
This is not surprising given the large ESMP2 doubles norms in these
states and the relatively low
%The norms of the first order wave functions for these states did follow the trend in errors, the $2^1$A$_1$ states had norms below 80\% while the other norms were above 83\% for the remaining states.
CC3 T\textsubscript{1} percentages of 85\% and 86\%
%for furan and pyrrole, respectively, while for the other states the values are above 90\%, making it likely that ESMP2's poor performance is due to the presence of doubly excited character.
%It is worth noting, however, that other wave function-based methods do not seem to have particularly large errors for these states. 
%Aside from the $2^1$A$_1$ states, for which the ESMP2
%errors are definitely the largest in magnitude,
%ESMP2 performs fairly well, being most comparable to the CC3.
As in many other molecules, regularization makes a big difference, and
$\epsilon$-ESMP2 reduces the errors in the $2^1$A$_1$ states to
below 0.05 eV while also lowering errors in most other states as well.
%Only the $3^1$A$_1$ state of furan was relatively untouched with the error changing from 0.26 eV to 0.25 eV.
%This immediately places $\epsilon$-ESMP2 as one of the most accurate methods presented here for these two molecules, and even performs better than MS-CASPT2 for pyrrole. 

% NOTE: I stepped away from many of the claims in the previous paragraph since the Rydberg issue may be making both forms of ESMP2 look spuriously good.

\subsubsection{Imidazole}

%$Another heterocyclic pentadiene to consider is imidazole, which contains two nitrogen atoms within the aromatic, five-membered ring.
%Particularly interesting due to its high reactivity and novel biological activity, this important building block is a motif in biological molecules
%with analgesic, antibacterial, antifungal, antiviral, antitubercular, and anticancer properties.\cite{HeterocyclesNaturalProducts_Pyrimidine_Imidazole} 
\begin{comment}
Here is what Roos says in their imidazole study, but they cite a book so I don't have a source I've read with this information -- 
''Imidazole is the active, functional group of histidine, one of
the 20 naturally occurring amino acids residues. Histidine can
act as a proton donor and acceptor, and, because of its low pK,
$\approx$6.2, it is partially protonated at neutral pH. Histidine is
frequently found as a ligand in metalloprotein complexes such
as carboxypeptidase A, azurin, myoglobin, and others. It also
often assumes a catalytic role such as the charge-transfer relay
system found in R-chymotrypsin and other serine proteases.''
\end{comment}

%Marine sponges 

Imidazole is a case where experimental comparison is particularly
challenging, as the UV-Vis spectrum has only been taken
in ethanol and aqueous solutions.
\cite{Roos_CASSCF_imidazole, heterocyclic_chemistry_book,raman_imidazole,bernarducci_imidazole,caswell_imidazole,fawcett_imidazole,gelus_heterocycles,grebow_imidazole}
Further, there is disagreement about whether
%As imidazole is highly polar, amphoteric, and partially protonated in neutral-pH solutions, the UV-Vis spectrum, which features two broad bands around 6.0 eV (CITATIONS Roos\_CASSCF\_imidazole refs 3-9,11 3-9) and 6.5 eV,(CITATIONS Roos\_CASSCF\_imidazole refs 3-9,11 5,8) is thus quite complex as both imidazole and its protonated ion imidazolium are present. 
%This leads to two competing theories --
imidazole and imidazolium (the protonated form)
have overlapping broad bands or each form separate strong peaks.
%spectra overlap into could both have two intense broad bands, or they could each only have a single strong peak.
\cite{Roos_CASSCF_imidazole} 
In any case, Thiel selected their CASPT2 results as the TBE
for three singlet vertical excitations:
the 1${}^1A$'' $n\rightarrow\pi^*$ state at 6.81 eV,
and two $\pi\rightarrow\pi^*$ excitations of A' symmetry
at 6.19 eV and 6.93 eV. 
%
%{\color{blue}Thiel's MS-CASPT2 energetics are more similar to experiment though, if we adopt the theory that the imidazole and imidazolium spectra both consist of two intense broad peaks that correlate to the 2 and 3 ${}^1A$' $\pi\rightarrow\pi^*$ transitions. 
%Let's consider the solvent effects though if we do adopt this theory. 
%In Thiel's MS-CASPT2 study, the 2 ${}^1A$' state is slightly less polar than the ground state, so polar solvents (e.g. ethanol) would not stabilize the excited state as much, leading to a blue shift (increase) in the excitation energy when imidazole is solvated. However, Thiel found this state to be 6.19 eV in the gas phase, which is larger than the peak of the experimental band at 6.00 eV, counter to what we expected. 
%Analysis of the 3 ${}^1A$' state does align with this logic, though. According to MS-CASPT2, the 3 ${}^1A$' state is more polar than the ground state, so solvation with a polar solvent like ethanol should lower the excitation energy, as we see from Thiel's gas-phase excitation energy of 6.93 eV versus the experimental band peak at 6.50 eV.
%Therefore, either the MS-CASPT2 excitation energy for the 2 ${}^1A$' state is high by at least 0.2 eV or the MS-CASPT2 dipole calculations for the ground state and the 2 ${}^1A$' state are erroneous.} 
%{\color{red}Can CASSCF dipole calculations be trusted? How quantitatively accurate are they? Side fun fact -- oftentimes, $\pi\rightarrow\pi^*$ excited states are more polar than the ground state according to Encyclopedia of Spectroscopy and Spectrometry 2017. }
%
CASSCF and CASPT2 calculations have shown that these states are not
Rydberg in nature. \cite{Roos_CASSCF_imidazole}

Interestingly, there is some disagreement between different CASSCF
approaches and also ESMF about the nature of these three states.
%although there are significant differences in the energies and natures of the CASSCF wave functions between Thiel's study, Roos's study, and our current CASSCF calculations. Thus it is likely that part of the difficulty of ESMF to converge to stationary points resembling these states arises from some lack of clarity of how these states should be described. 
%
If we consider the 6-electron, 5-orbital $\pi$ system in imidazole,
we would not expect the lowest-energy and nodeless 1a'' $\pi$ orbital
to participate strongly in low-lying excitations.
Instead, the
%As the lowest-energy $\pi$ orbital (referred to further as the 1a'' orbital) that has no nodes other than that of the molecular plane is relatively low in energy relative versus the
occupied 2a'' and 3a'' $\pi$ orbitals,
which both have an additional nodal plane, 
%we would not expect the 1a'' orbital to participate significantly in the lowest few valence excitations of A' symmetry. Neglecting it, this leaves us with a 4-orbital $\pi$ system consisting of the occupied 2a'' and 3a'' $\pi$ orbitals and
can form up to four singlet excitations into the
unoccupied 4a'' and 5a'' $\pi^*$ orbitals.
%Linear combinations of these four orbitals thus imply the existence of four $\pi\rightarrow\pi^*$ excitations concerning these orbitals.
%
Thiel's work on MS-CASPT2 and CC3 shows the $2^1$A$^\prime$
state as being dominated by the 3a''$\rightarrow$4a'' excitation,
while the $3^1$A$^\prime$ state involves a positive superposition
of the 2a''$\rightarrow$4a'' and 3a''$\rightarrow$5a'' excitations.\cite{CC_Thiel_benchmark} 
However, Roos found the states of A$^\prime$ symmetry to both
have significant contributions from each of the
3a''$\rightarrow$4a'', 2a''$\rightarrow$4a'',
and 3a''$\rightarrow$5a'' excitations.
\cite{Roos_CASSCF_imidazole} 
In our ESMF results, the A$^\prime$ states are essentially
the plus and minus combinations of the 3a''$\rightarrow$4a'' and
3a''$\rightarrow$5a'' transitions, with very little
contribution from 2a''$\rightarrow$4a''.
Taken together, these results show that imidazole is a case
where the exact mixing of the components within excited
states is quite sensitive to the amount of correlation
and orbital relaxation in play.
%% So this section about solvents is a little out of place but I was thinking about why we would adopt one CASSCF study versus another, and that imidazole is solvated in the UV-Vis spectrum was bothering me when I looked at Thiel's MS-CASPT2 calculations. Roos actually has the opposite problem though with dipoles. Both of Roos's numbers are too high, and state 2 has a higher dipole than the GS while state 3 has a lower dipole than the GS. I am not offended if we take this out, just thought it was interesting. 

%For ESMF, we have actually found two states for the 2${}^1A$' state -- one that looks like Thiel's MS-CASPT2 result and is strongly characterized by the 3a''$\rightarrow$4a'' excitation, 
%and another that resembles Roos's and our CAS results and is dominated by a mixture of the 2a''$\rightarrow$4a'' and 3a''$\rightarrow$5a'' excitations. In the former, the ESMP2 method errors low versus Thiel's MS-CASPT2 energy by only -0.04 eV, while in the latter, ESMP2 errors by -0.54 eV versus the best estimate.

In terms of energetics, ESMP2's excitation energies have an
overall accuracy similar to that of CC3.
Regularization only improves the accuracy in one of the three states,
making this an unusual molecule in that regard and raising the
question of how accurately ESMF has captured the zeroth order
representation, especially in light of the disagreement between
it and multiple versions of CASSCF in the A$^\prime$ states.
It seems possible that this is a case where the primary
singly excited components are close enough in energy that
how they mix is substantially affected by correlation effects
from doubly excited determinants, which is an effect that
is simply beyond the reach of ESMF.

\subsubsection{Pyridine}
%An excellent solvent and a common starting material in pharmaceuticals and agrochemicals, pyridine is produced on large scales, at more than 20,000 tons annually.\cite{UllmansEncyclopedia_Pyridine} A motif in biologically active compounds including nicotine, vitamin B${}_3$, and vitamin B${}_6$, this mildly basic, $\pi$-electron-deficient azabenzene is more common in natural products in its aromatic form than in any of its other oxidative states.\cite{HeterocyclesNaturalProducts_Pyridine,UllmansEncyclopedia_Pyridine}

We studied the 1 ${}^1 B_1$ and 2 ${}^1 A_2$ $n\rightarrow\pi^{*}$ excitations and four $\pi\rightarrow\pi^*$ states: 1${}^1 B_2$, 2${}^1 A_1$, 3${}^1 A_1$ and 2${}^1 B_2$.
The best estimates for these states comes from Nakatsuji et al's SAC-CI calculations, \cite{Nakatsuji2001_pyridine_SAC-CI} which are close to experimental gas-phase excitation values. Using ESMP2 to predict the excitation energies for these states led to a mixture of errors. $2^1$A$_1$ and $2^1$A$_2$ had the smallest errors of 0.04 and 0.14 eV, respectively. $1^1$B$_2$ and $1^1$B$_1$ had errors of 0.33 and 0.34 eV, and the largest errors were from the $3^1$A$_1$ and $2^1$B$_2$ states with 0.63 and 0.73 eV.
None of these states show especially large doubles norms, at least not
compared for example to those seen in the polyenes.
%The singles norms only somewhat correlate with the magnitude of the errors.
%Both states with errors above 0.5 eV had norms of only 80\% while the two states with norms below 0.2 eV had norms of 83/84\%. However the smallest norm and largest norm were for the $1^1$B$_2$ and $1^1$B$_1$ states, respectively, even though these states produced similar errors that were not extremely small or large.
%The states of pyridine all generally have doubly excited character (as stated in the Wiberg TD-DFT benchmark) so somewhat high errors are not unexpected, but the magnitude of the errors do not correspond completely with the singles norms from the ESMP first order wave function or the CC3 T\textsubscript{1} percentages. However, the largest errors coming from the $3^1$A$_1$ and $2^1$B$_2$ states can be at least partially explained by looking at the calculated overlaps: the $3^1$A$_1$ state overlaps with the ground state and the $2^1$B$_2$ state overlaps with the $1^1$B$_2$ state.
Although ESMF and EOM-CCSD both agree that the excited states of A$_1$ symmetry
are superpositions of two main components, ESMF predicts much more
equal superpositions than EOM-CCSD.
%The predicted CI vector from ESMF also predicts a much more equal weighting between the two main single excitations that describe the
%state than EOM-CCSD does. 
Compared to other methods, ESMP2 is overall slightly more
accurate than the coupled cluster methods, although the accuracy
of both varies significantly from state to state.
The overall accuracy of $\epsilon$-ESMP2 is better, but this
comes from improvements in some states partially counteracted
by detriments in others.
%does similarly to Thiel's CASPT2, and worse than SA-CASSCF.
%Oddly adding the level shift to ESMP2 actually increased the errors all but the $3^1$A$_1$ and $2^1$B$_2$ states, whose errors were dramatically decreased to below 0.15 eV.
%The two highest accuracy ESMP2 results (states $2^1$A$_1$ and $2^1$A$_2$) had the largest increase in error after adding the level shift. However, even after these changes, $\epsilon$-ESMP2 is still one of the more accurate methods shown, only SA-CASPT2 does slightly better on average. 

%In pyridine there are two $n\rightarrow\pi^{*}$ excitations that are studied, 1 ${}^1 B_1$ and 2 ${}^1 A_2$, both had ESMP2 errors under 0.4 when compared to the Thiel best estimate. The two lower energy $\pi\rightarrow\pi^*$ states, 1 ${}^1 B_2$  and 2 ${}^1 A_1$ had similarly small errors while the two higher energy $\pi\rightarrow\pi^*$ states, 3 ${}^1 A_1$ and 2 ${}^1 B_2$ had errors slightly above 0.5 eV. These slightly larger errors are not surprising as the EOM-CCSD results reported in the Thiel paper showed that these two states had larger contributions from excitations beyond singles than the other four states studied here and our method can only account for contributions from single excitations. From our overlap analysis, the ESMF wave functions for these two states also had non-negligible overlap with other excited states. For $1^1$B${}_2$ there was a large overlap with $1^1$B${}_2$ and for $3^1$A${}_1$ there was a slightly smaller, but still non-negligible, overlap with the Aufbau determinant. See table (label) in the SI. 

\subsubsection{Pyrazine}
%https://www.sciencedirect.com/science/article/pii/B9780444828880500398
%Pyrazine is commonly used to create fluorescent dyes for biological experimentation and energy transfer materials. 

For pyrazine, four $n\rightarrow\pi^*$ states -- $1^1$B${}_{3u}$, $1^1$A${}_u$, $1^1$B${}_{2g}$, $1^1$B${}_{1g}$ -- and four $\pi\rightarrow\pi^*$ states -- $1^1$B${}_{2u}$, $1^1$B${}_{1u}$, $2^1$B${}_{1u}$, $2^1$B${}_{2u}$ -- were studied. 
Thiel selected EOM-CCSD($\widetilde{T}$) calculations
for the TBEs, \cite{Thiel-BM-MAIN}
%to define their best estimates, these values match fairly well to available experimental data for the states that have experimental estimates.
compared to which ESMP2 produced a wide variety of errors, ranging
from 0.10 eV in the $1^1$B$_{1u}$ state to 1.18 eV in the $1^1$B$_{1g}$ state.
%This information matches the trend shown by the singles norms, the lowest error corresponded to the largest singles norm of 83\% while the highest error corresponded to the smallest singles norm for this molecule with 77\%.
%As this is a heterocyclic aromatic molecule, many of the studied states have character from higher excitations which ESMF cannot describe, partially causing the errors seen here.
%The errors of the $1^1$B$_{2u}$ and $2^1$B$_{2u}$ states might have also been impacted by a nonzero overlap between them.
$\epsilon$-ESMP2 shows significant improvements,
with errors of less than 0.25 eV in all states. 
%Looking at the T\textsubscript{1} percentages from the CC3 calculations reported in the 2008 Thiel benchmark, all the studies pyrazine states are below 90\%, with the exception of the $1^1$B$_{1u}$ and $2^1$B$_{1u}$ states. It is unclear why the $2^1$B$_{1u}$ state has a significantly larger error than the $1^1$B$_{1u}$ state.
The CC methods do better than ESMP2 but worse than $\epsilon$-ESMP2.
The same is true of CASPT2, although its accuracy is much closer to
that of $\epsilon$-ESMP2.
%For the other methods studied in the original benchmarking paper, these methods generally do better than ESMP2 with no level-shift, EOM-CCSD does error higher on average, but has a much smaller range of errors than ESMP2.
%When the aggressive level shift of 0.5 Ha is added the accuracy of the method becomes comparable to the higher scaling MS-CASPT2 method and becomes more accurate than the coupled cluster methods. 

\subsubsection{Pyrimidine}
%natural product\cite{HeterocyclesNaturalProducts_Pyrimidine_Imidazole} Need to look at this before using it

%https://reader.elsevier.com/reader/sd/pii/B9780128175927000095?token=9327CA6A03BBD425A2F33DECCC81CE2B379A4F91E8A33B97656B01A42BF54DDFE9606BDDF98B674DFC0445D45A7FF0BD&originRegion=us-east-1&originCreation=20210812020742
%Pyrimidine is a compound that is ubiquitous in organic synthesis. It is a building block or decomposes into the nucleobases and is a key part of pharmaceutical compounds. 

Four excited states were studied for pyrimidine, two $n\rightarrow \pi^*$ states -- $1^1$B$_{1}$ and $1^1$A$_2$ -- and two $\pi\rightarrow\pi^*$ states -- $1^1$B$_{2}$ and $2^1$A$_1$.
The Thiel best estimates for the excited states of pyrimidine were based on coupled cluster results with non-iterative triples and basis set corrections. Generally these values error a few tenths of an electron volt high compared to experimental values. Based on the work presented in benchmarking studies by Loos, et. al.\cite{mountaineering_medium_molecules} and Schreiber et. al.\cite{Thiel-BM-MAIN} most ab initio methods error high for these states compared to experiment, not only the ones based on coupled cluster.
An unusual feature in pyrimidine as compared to the other azabenzenes
studied here is that EOM-CCSD does comparably to Thiel's CASPT2
and only slightly worse than CC2 and CC3.
In the other azabenzenes, EOM-CCSD had noticeably higher errors
when compared to CC2 and CC3.
%For pyrimidine the two $n\rightarrow\pi^*$ excitations, $1^1$B${}_1$ and $1^1$A${}_2$ had markedly higher errors than the two $\pi\rightarrow\pi^*$ states, $1^1$B${}_2$ and $2^1$A${}_1$, though this is not a pattern seen in the other azabenzenes. This molecule presents one of the few cases where the higher energy excitations had lower errors when compared to the Thiel best estimates. Also unusual is that the norms calculated from the first order wave function amplitudes do not follow the same trend as the errors. the $2^1$A$_1$ and $1^1$B$_2$ states have (albeit slightly) smaller norms than the other two states despite having the smaller errors. All states have errors close to or above 0.5 eV, so it is possible that the magnitude of these norms just shows that ESMP2 will be inaccurate for all four states due to the presence of doubly or higher excited character. The CC3 T\textsubscript{1} values confirm this theory. The $2^1$A${}_1$ state also has a nonzero overlap with the ground state which could be contributing to the magnitude of the error. 
Without a level shift, ESMP2 is most similar in accuracy to Roos's CASPT2, with
both methods producing large errors.
$\epsilon$-ESMP2 shows smaller errors, ranging from 0.08 to 0.31 eV,
which is closer to but not as accurate as Thiel's CASPT2 and the CC methods.
%though $\epsilon$-ESMP2 is still slightly less accurate than these four methods. 

\subsubsection{Pyridazine}

Three $n\rightarrow\pi^*$ states were studied in pyridazine: $1^1$B${}_1$, $1^1$A${}_2$, and $2^1$A${}_2$.
While all of these states had ESMP2 errors above 0.5 eV,
those for the $2^1$A${}_2$ and $2^1$A${}_1$ states
were particularly large at almost 1 eV.
%and a non-negligible overlap with $2^1$A${}_2$.
%The one $\pi\rightarrow\pi^*$ state studied here, $2^1$A${}_1$, had a similarly large error of 0.90 eV when compared to the Thiel best estimate for the state. 
%The singles norms are all around 80\% for these states, showing that there likely is doubly and higher excited character in the description of all fours states. 
Pyridazine is thus another good example of ESMP2's difficulties
in larger $\pi$ systems, but it is also one of the most powerful
examples of the practical efficacy of regularization,
with all of $\epsilon$-ESMP2's errors coming in at less than 0.25 eV.
%While these relatively large errors are disappointing, it is worth pointing out that this molecule presents one of the best examples of the impact the 0.5 Ha level shift can have on the ESMP2 results. Adding the level shift reduces these errors which were all above 0.5 eV to 0.11, 0.07, 0.02, and 0.24 for $1^1$B$_1$, $1^1$A$_2$, $2^1$A$_1$, and $2^1$A$_2$, respectively.
Comparing these results to other wave function methods, ESMP2 was
easily the least accurate, while $\epsilon$-ESMP2 performed similarly
to CC2, was more accurate than EOM-CCSD, and was only slightly less
accurate than CC3.
The TBEs for this molecule were directly taken from Thiel's
CASPT2 values, so it is difficult to make a fair comparison to
that method.

\subsubsection{\textit{s}-Triazine}
%Triazine is an incredibly useful molecule in organic synthesis as it can be used in place of HCN in many applications. It is also well known industrially as being a key ingredient for several types of pesticides.\cite{triazine_as_HCN, triazine_pesticide}

In triazine we studied three $n\rightarrow\pi^{*}$ states --  $1^1$A$^{\prime\prime}_1$, $1^1$A$^{\prime\prime}_2$, $1^1$E$^{\prime\prime}$ -- and one $\pi\rightarrow\pi^{*}$ state, $1^1$A$^\prime_2$.
Even compared to its performance on other azabenzenes,
ESMP2 did poorly here with typical errors around 1 eV,
making it by far the least accurate among the wave function methods.
The CC3 T\textsubscript{1} values are all below 90\%
and the ESMP2 doubles norms all above 0.35, suggesting that triazine
is simply a particularly painful example of the difficulty
unregularized ESMP2 has when an extended $\pi$ system brings
the lowest doubly excited configurations too close to the
primary singly excited configurations.
%However percentages of similar magnitudes are seen in other azabenzene states that have errors closer to 0.75 eV than the over 1 eV errors seen for triazine. The norms calculated from the first order wave function amplitudes are correspondingly small, all fall in the range of 76-79\%. 
%Unfortunately, the variety of sources for TBEs among the heterocycles
%--- Triazine, tetrazine, pyridazine, and imidazole use Thiel's CASPT2 data,
%while furan, pyrrole, pyrimidine and pyrazine use CC calculations and
%pyridine uses SAC-CI ---
%makes it difficult to make any systematic statements about why
%triazine might be expected to produce particularly difficulty for ESMP2.
%When looking at the experimental results from a 1984 absorption study
%$\textcolor{red}{(citation)}$
%of the azabenzenes that was used to help justify TBE choices,
%it is worth noting that the experimental values for the
%$2^1$A$^{\prime}_2$ and $1^1$E$^{\prime\prime}$ states differ
%greatly from their TBEs.
%In particular, the $1^1$E$^{\prime\prime}$ experimental value
%is recorded as 3.97 eV while the TBE value is 4.71 eV.
%If the experimental value is trusted, our method performs oddly well for this state as we predict an energy of 3.49 eV. However, the $1^1$E$^{\prime\prime}$ state is known to be incredibly difficult to study as it has a known degeneracy with the $1^1$A$^\prime_2$ state that makes the "analysis of this band system extremely complicated."\cite{azabenzene_experiment}
Again, $\epsilon$-ESMP2 significantly mitigates this difficulty,
reducing the worst error to
%If we instead look at the results of using a 0.5 Ha level shift with ESMP2 our triazine results, as with the other molecules studied, our method becomes much more promising. Our errors are reduced by around 1 eV with the largest error for this molecule now at
0.31 eV instead of 1.45 eV.
$\epsilon$-ESMP2 is still less accurate than CC2 and CC3, but is
comparable to EOM-CCSD and noticeably better than Roos's CASPT2. 

%The ESMP2 results for this molecule were all above 1 eV, very unlike the ESMP2 results for the other azabenzenes studied. This difference is unusual because these states did not have more doubly excited character than the states studied in the other azabenzene molecules, none had large overlaps with other states, convergence in the ESMF algorithm was not unusually difficult for any of these states, and the ESMP2 amplitudes were comparable to those of the other azabenzene results. Further study will need to be done to determine the source of this uniform large error. We purposefully did not study the doubly excited state presented in Thiel's original benchmarking paper here as the ESMF ansatz cannot capture doubly excited character. 

% will add in citations later
\subsubsection{\textit{s}-Tetrazine}
%Tetrazine is an important tool in organic synthesis as it is often used as a basis for reagents in Diels-Alder reactions to form substituted pyridazine rings. These types of reactions and the fluorescence properties of tetrazine have caused tetrazine derivatives to become widely used for biorthogonal labeling. % Flourescent heterocycles: Recent trends and new developments. Schramm, Weiss. Chapter 2 of Advances in Heterocyclic Chemistry. 2019

In this molecule we look at four $n\rightarrow\pi^*$ excitations -- $1^1$B$_{3u}$, $1^1$B$_{1g}$, $1^1$B$_{2g}$, and $2^1$A$_{u}$ -- and two $\pi\rightarrow\pi^*$ excitations -- $1^1$A$_u$ and $1^1$B$_{2u}$.
Note that we have not studied the strongly doubly excited
%As our ESMF wave function ansatz makes the method unsuited for studying doubly excited states we did not study the
$1^1$B$_{3g}$ state, and indeed the original Thiel
benchmark does not even contain CC numbers for this state. 
% Need to write up my analysis of the experimental results to say whether that choice for the TBE's is okay or not
% used the experimental studies from both the Thiel and Mountaineering papers, will cite later
As in a number of other heterocycles, Thiel's TBEs for tetrazine were taken
directly from the CASPT2(b) results without basis set extrapolation.
%There are many arguments against using CC2 data for the TBEs, even with a large basis set that can properly account for Rydberg character, however we will still used these values for the sake of consistency.
%Looking at a few experimental studies, these best estimates match for the states of symmetry ungerade symmetry. Experimental values for the excitation energies of the states of gerade symmetry were not found. 
%Of the methods studied in the original benchmark,\cite{Thiel-BM-MAIN}
Against this TBE, CC2 and CC3 show errors mostly below about 0.3 eV,
whereas EOM-CCSD errors are higher.
As in triazine, all of tetrazine's CC3 \%T$_1$ values for the states studied
are between 80 and 90, again implying a difficult
playing field for unregularized EMSP2,
%These percentages matched the information gained from the norms of the first order wave functions for each state, all were between 76\% and 78\%, relatively small for a molecule of this size. 
which duly makes errors on the order of
%Our non-level-shifted ESMP2 results performed similarly to those of triazine, with errors generally around
1 eV and shows doubles norms above 0.35.
%, making ESMP2 the least accurate out of the methods studied here.
ESMP2 was especially bad for the $2^1$A$_{u}$ state,
with an error of 1.86 eV, its worst error among the heterocycles.
%which was much higher than the errors of any other states studied in this group.
%The generally high errors for these states can be attributed to the non-negligible doubly excited character in all of these states. Since our reference method only includes information about single excitations, even small amounts of doubly excited character can be an issue.
Introducing regularization via $\epsilon$-ESMP2 makes a huge difference,
%When we add in an aggressive level shift of 0.5 Ha the errors decrease dramatically for all states.
reducing errors to less than 0.22 eV in all states except for $2^1$A$_u$.
The $2^1$A$_u$ state's error only falls to 0.72 eV,
making it one of the worst for $\epsilon$-ESMP2, especially considering
that the ESMP2 doubles norm, although not small at 0.41, is
not as large as in the difficult benzene or polyene states
and so it is not as clear in this case that ESMP2 would be able
to predict $\epsilon$-ESMP2's failure.
%still a dramatic reduction, but not as small as what we were expecting given the level shifted data we collected for similar molecules. Looking at our overlap data, this remaining large error may be due to a non-negligible overlap between the $2^1$A$_u$ state and the $1^1$A$_u$ state.
Outside of the $2^1$A$_u$ state, the accuracy of $\epsilon$-ESMP2 is
similar to the other wave function based methods.
%A direct comparison with MS-CASPT2 cannot be done as the TBEs for this molecule are the MS-CAPST2 results.  

\subsection{Group 5: Nucleobases}

This group of molecules includes cytosine, thymine, uracil, and adenine. Due to the size of these molecules, there were no CC3/TZVP calculations
reported in the original Thiel benchmark, \cite{Thiel-BM-MAIN} and
we have instead taken the CC3 data from a more recent study.
\cite{kannar2014-nucleobases}
Thiel selected CC2/aug-cc-pVTZ results for the TBEs in
all nucleobase states. 
We note that many of the nucleobase states were
difficult for ESMF to converge to, which was even true in some
cases in which the state was dominated by a single singly excited
component.
In both of the cases where ESMF failed to converge to a stationary point
%some of these states were very well described by a single CSF, so it is currently unclear why the difficulties occurred. These included
(the $2^1$A$^\prime$ state of cytosine and the $3^1$A$^\prime$ state of uracil),
we hypothesize that a loss of good orthogonality with lower states
during the ESMF optimization was partially to blame.
Still, it is not clear why this was such an issue in these cases,
as at least a small loss of orthogonality is normal in ESMF due to
its state-specific orbital relaxation, and the same difficulty was not
present in most other states.
%. As these states are both of symmetry A$^\prime$ we hypothesize that a lack of orthogonality between the ground and excited states may have contributed to the issue. A third state, also of $A^\prime$ symmetry will also not be included in any further analysis based on MUEs or standard deviations, the$3^1$A$^\prime$ state of thymine had a singles norm below 70\% showing that the ESMP2 result should not be used.  

For the excited states in this group, ESMP2 had an MUE of 0.61 eV,
making it the wave function method with the worst overall accuracy,
as seen in Table \ref{tab:wfn_red_gray}.
%on average the least accurate method for predicting the excitation energies of these states. The standard deviation was calculated to be 0.46 eV, making ESMP2 also have the highest spread of errors compared to the other methods.
$\epsilon$-ESMP2, on the other hand, had a much smaller MUE of 0.19 eV,
%and a standard deviation of 0.15 eV,
putting it on par with CC2 and ahead of EOM-CCSD.
%making it more accurate than all of the other methods shown here except for MS-CASPT2. The spread of errors from CC2 was smaller, however the MUE for CC2 was 0.02 eV higher than that of $\epsilon$-ESMP2. 
%As in other groups of molecules, ROKS
%study\cite{VanVoorhis2013_ROKS} on the lowest excited states of these molecules showed once again that basis set size plays a very important role for the prediction of excited state energies with this method. Changing the basis set from 6-31G* to 6-311+G* decreased the predicted energy anywhere from less than 0.1 eV to almost 0.7 eV, depending on the molecule.
%However, unlike the other molecule groupings, switching from the RO-PBE0 functional to RO-LC$\omega$PBE0 had a much larger impact on the energies, changing them by over 0.2 eV, though whether it resulted in a decrease or increase in energy was dependent on the molecule.
$\epsilon$-ESMP2 was also more accurate in the nucleobases
than the TD-DFT methods shown in Table \ref{tab:td_dft_red_gray},
while being a little less accurate than DFT/MRCI.

\subsubsection{Cytosine}
% Thiel doesn't cite any experimental work. I could not find any experimental papers for gas phase cytosine on it's own. Many were looking at the stacked pairs or the pairs in solution. 
% other papers jacki listed (not bartlett) seem to be calculations done on the paired nucleobases, so not a direct comparison to what we (or Thiel) did

In cytosine, we studied two $\pi\rightarrow\pi^*$ states -- $2^1$A$^\prime$ and $3^1$A$^\prime$ -- and two $n\rightarrow\pi^*$ states -- $1^1$A$^{\prime\prime}$ and $2^1$A$^{\prime\prime}$.
%A 2012 benchmark for these states that looked at a few high-level coupled cluster methods found estimates for the excited state energies that were similar to the TBEs. Not all the data reported in this paper were using the same basis set and geometry, making it impossible to do a direct comparison between the different calculated energies. However, it does seem that the methods agreed on the TBE for the $2^1$A$^\prime$ and $3^1$A$^\prime$ states while for the other states there is more variation. It is also important to note that the $1^1$A$^{\prime\prime}$ state has a great deal of Rydberg character, which is why there is such a discrepancy between the different calculation types and basis sets for this state. For the $2^1$A$^{\prime\prime}$ state it is less clear why there is a large discrepancy between excitation energies between these different methods.\cite{Bartlett2012_nucleobases} So while there are arguments to whether CC2 results should have been used as the best estimate for these states, we will still use these best estimates in our analysis of ESMP2 for the sake of consistency. 
%Cytosine was unable to converge to a stationary point resembling the lowest energy excited state, the$2^1$A$^\prime$ state, despite the state being dominated by a single HOMO$\rightarrow$LUMO excitation.  We are currently unsure why ESMF did not converge to this state, however it might be due to orthogonality issues since this state is of the same symmetry of the ground state.
In the states ESMF successfully converged, ESMP2 with no level shift
did fairly poorly with errors between 0.5 and 1 eV, which is similarly
inaccurate to
%Out of the methods presented in the 2008 benchmarking paper ESMP2 performs most similarly to
EOM-CCSD and worse than the other wave function methods.
%The singles norms do not provide much insight into this poor accuracy as the norms for the $1^1$A$^{\prime\prime}$ and $2^1$A$^{\prime\prime}$ state are at 79\% and 82\% respectively, which seems somewhat typical for a molecule of this size, with errors above 0.2 eV lower percentages would have been expected. The larger error of 0.84 eV from the $3^1$A$^\prime$ state does correspond to a markedly lower norm percentage of 76\%.
ESMP2 shows a notably peculiar result for the $2^1$A$^{\prime\prime}$
state in that it overestimates the excitation energy.
In most other cases, ESMP2 tends to error low.
%energy and its first order wave function amplitudes are relatively small. While the weighting of the different contributions to the ESMF wave function for this state does match what EOM-CCSD predicted, it is possible that EOM-CCSD gave a poor description for the state and ESMF did not actually converge to the $2^1$A$^{\prime\prime}$ state.
This $2^1$A$^{\prime\prime}$ state remains an outlier even after
introducing regularization, with the error barely changing.
%decreased and the ESMP2 continues to overestimate the energy
In contrast, regularization brings the errors for the other two
below 0.2 eV.
%Another possibility from this unusual behavior is that CC2, the method used for the TBE for these states, was less accurate for this state than the others. 
%Aside from the $2^1$A$^{\prime\prime}$ state, adding the level shift made $\epsilon$-ESMP2 extremely accurate for these excited states, even rivaling MS-CASPT2 results. 

\subsubsection{Uracil}
% not included in bartlett's paper, see https://pubs.acs.org/doi/pdf/10.1021/ct500495n or https://link.springer.com/content/pdf/10.1007/s00894-014-2503-2.pdf, will add these to the bibliography later

%For uracil we studied three $n\rightarrow\pi^*$ states -- $1^1$A$^{\prime\prime}$, $2^1$A$^{\prime\prime}$, and  $3^1$A$^{\prime\prime}$ -- and three $\pi\rightarrow\pi^*$ excitations -- $2^1$A$^{\prime}$, $3^1$A$^{\prime}$, and $4^1$A$^{\prime}$.

%Like in cytosine, the TBEs for uracil come from CC2 results with an augmented basis, however in cytosine this was an aug-cc-pVTZ basis while for uracil it is an aug-cc-pVQZ basis. Looking at high level coupled cluster results (will add citations to papers linked in comments above) the TBE values could be fairly accurate for these states. CCSD(T) and CCSDR(3) with the TZVP basis set predicted excitation energies a around 0.2 eV larger for these states, however going from the TZVP basis set to an augmented basis set has been shown to usually lower the excitation energy by a few tenths of an electron volt, cancelling out the effect. So we will be using the TBEs presented in the 2008 benchmarking paper to determine the accuracy of ESMP2 in uracil as they generally agree with the results from high-level ab initio methods. 

Five states were successfully studied in uracil: the $n\rightarrow\pi^*$ states with symmetry labels $1^1$A$^{\prime\prime}$, $2^1$A$^{\prime\prime}$, and $3^1$A$^{\prime\prime}$ and the $\pi\rightarrow\pi^*$ states with symmetry labels $2^1$A$^\prime$ and $4^1$A$^\prime$.
ESMF failed to converge the $3^1$A$^\prime$ state in uracil.
Similar to the cytosine state, this state is dominated by a single
singly excited component, so the reasons for this failure are not
obvious.
%HOMO-1$\rightarrow$LUMO excitation and had the same symmetry of the ground state.
%It is possible that the lack of orthogonality considerations currently built into our optimization scheme for ESMF is the cause of this failure.
In the other states, ESMP2 does very well for the $1^1$A$^{\prime\prime}$
state with an error of just 0.02 eV, somewhat poorly for the
$2^1$A$^\prime$ and $2^1$A$^{\prime\prime}$ states with errors around 0.5 eV,
and very poorly for the $3^1$A$^{\prime\prime}$ and $4^1$A$^\prime$ states.
%The singles norms did generally correspond to the magnitude of the error with the $3^1$A$^\prime\prime$ and $4^1$A$^\prime$ states having norms of 74\% while the other states had norms of 79\% or above. This means that the relatively high errors for the $3^1$A$^\prime\prime$ and $4^1$A$^\prime$ states are likely due to the presence of doubly excited character. In the $4^1$A$^\prime$ state we also had a nonzero overlap with the ground state, which could be contributing to the magnitude of the error.
Overall, ESMP2 performs comparably to EOM-CCSD, but worse than the other wave function methods.
Regularization reduces error for the $2^1$A$^\prime$ state to just 0.01 eV,
leaves the error in the $1^1$A$^{\prime\prime}$ state essentially unchanged,
and brings the other states' errors to around 0.3 eV.
This places the accuracy of $\epsilon$-ESMP2 ahead of EOM-CCSD and CC2,
but still behind that of Thiel's CASPT2. 

% thymine 0.5 Ha level shift calculation hit the time limit so I'm still waiting on two states. 
\subsubsection{Thymine}
For thymine we studied two $n\rightarrow\pi^*$ states, $1^1$A$^{\prime\prime}$ and $2^1$A$^{\prime\prime}$, and three $\pi\rightarrow\pi^*$ states -- $2^1$A$^\prime$, $3^1$A$^\prime$, and $4^1$A$^\prime$. 
ESMP2 does very well for $1^1$A$^{\prime\prime}$,
achieves errors of around 0.5 eV for $2^1$A$^\prime$ and $2^1$A$^{\prime\prime}$, and has errors of over 1.5 eV for $3^1$A$^\prime$ and $4^1$A$^\prime$.
%If the results for the $3^1$A$^\prime$ and $4^1$A$^\prime$ states are neglected, ESMP2 performs similarly to EOM-CCSD but worse than the other methods presented in the 2008 benchmarking paper.
For both of the latter states, large doubles norms of 0.497 and 0.476, respectively, warn of the trouble.
%$3^1$A$^\prime$ and $4^1$A$^\prime$ have extremely large errors compared to any of the results presented in the original paper.
%However, looking at the singles norms for these states, the $3^1$A$\prime$ state has a norm of 69\%, which falls below our cutoff for the inclusion in the statistical analysis. The $4^1$A$^\prime$ state has a norm of 70\%, just above the cutoff but still extremely low for molecules of this size.
%CC3 calculations reveal that these states have below 90\% singles character, and based on the sensitivity our method has to the presence of doubly excited or higher character, the errors of around 0.5 eV were expected.
%The reason why the $3^1$A$^\prime$ and $4^1$A$^\prime$ have errors above 1.5 eV likely arises partially from the lack of orthogonality between these states. Both states have nonzero overlap with the ground state, the $2^1$A$^\prime$ state and an overlap of 67\% between each other. 
$\epsilon$-ESMP2 has much lower errors for all states and makes
particularly large improvements in the $3^1$A$^{\prime\prime}$ and
$4^1$A$^\prime$ states, with an overall accuracy in this molecule
%While adding the level shift cannot correct any orthogonality issues as they stem from the reference method, from these results it seems as though $\epsilon$-ESMP2 can more correctly simulate the presence of higher excited character in these states.
better than EOM-CCSD or Roos's CASPT2 but worse than CC2 and
Thiel's CASPT2.

\subsubsection{Adenine}

For adenine we studied two $\pi\rightarrow\pi^*$ states, $2^1$A$^\prime$ and $3^1$A$^\prime$, and two $n\rightarrow\pi^*$ states, $1^1$A$^{\prime\prime}$ and $2^1$A$^{\prime\prime}$. 
ESMP2 gives errors above 0.5 eV for $2^1$A$^\prime$ and $2^1$A$^{\prime\prime}$ and errors of around 0.3 eV for the $3^1$A$^\prime$ and $1^1$A$^{\prime\prime}$ states,
%The smallest singles norm does correspond to the state ($2^1$A$\prime$) with the largest error of 0.84 eV, however the $2^1$A$^{\prime\prime}$ state also has a relatively large error of 0.63 eV but a norm of 74\% when a smaller norm would have been expected based on the other adenine and naphthalene results.
%It is possible that the large error is mostly due to nonzero overlaps between $2^1$A$^\prime$, $3^1$A$^\prime$ and the ground state and between $2^1$A$^{\prime\prime}$ and $1^1$A$^{\prime\prime}$. The norm of the singles amplitudes does not directly depend on this, thus the relatively high singles percentage for a molecule of this size in the case of the $2^1$A$^{\prime\prime}$ state.
with an overall accuracy comparable to EOM-CCSD but worse than the
other wave function methods.
$\epsilon$-ESMP2's worst error was 0.32 eV for the $1^1$A$^{\prime\prime}$
state, with its other errors all around 0.1 eV, making
it much more comparable to CC2, although not as accurate as Thiel's CASPT2.

\section{Conclusion}

We have applied ESMP2 and its regularized $\epsilon$-ESMP2 cousin
to the singlet excitations in the 28 molecules of the Thiel set,
which has clarified multiple aspects of this excited-state-specific
perturbation theory's behavior and performance.
First, we found that the underlying ESMF possesses a well-defined
excited-state-specific stationary point in 100 out of the 103 states
tested, suggesting that such stationary points typically exist
for singly excited singlet states in single-reference molecules.
Second, we found that ESMP2 is highly sensitive to the size of a molecule's
$\pi$ system. %, much more than is typical for traditional LR methods.
For molecules with five or fewer orbitals in their $\pi$ systems,
unregularized ESMP2's mean unsigned error was 0.32 eV, while for
molecules with six or more orbitals in their $\pi$ system it was 0.71 eV.
Third, this sensitivity closely tracks the size of the ESMP2
doubles norm, which helps us understand the issue as a straightforward
failure of perturbation theory brought about by
doubly excited configurations that are too close in energy to the
primary singles in the zeroth order reference.
%\textcolor{red}{as this reference operates
%under the assumption that doubly excited determinants would be small.}
Fourth, although this sensitivity is bad news for accuracy, it allows
the unregularized ESMP2 doubles norm to act as a reasonably effective
predictor of doubly excited character.
Finally, this sensitivity can be mitigated by repartitioning the
zeroth order approximation via a level shift, resulting in the
regularized $\epsilon$-ESMP2 method that outperforms TD-DFT, CC2, EOM-CCSD,
and even CC3 in overall accuracy on the singlet states in the Thiel set.
While CASPT2 showed the highest overall accuracy, $\epsilon$-ESMP2's
unsigned error of just 0.17 eV on singly excited states
was the lowest among methods that do not rely on an active space.

\section*{Supplementary Material}
See \href{run:SI.tex}{supplementary material} for more detailed information of the electronic structure calculations presented in this article.

\section{Acknowledgements}
This work was supported by the National Science Foundation's
CAREER program under Award Number 1848012.
Calculations were performed using the Berkeley Research Computing Savio cluster, the Lawrence Berkeley National Lab Lawrencium cluster, and the National Energy Research Scientific Computing Center, a DOE Office of Science User Facility supported by the Office of Science of the U.S. Department of Energy under Contract No. DE-AC02-05CH11231.
J.A.R.S. and H.T. acknowledge that this material is based upon work
supported by the National Science Foundation Graduate Research
Fellowship Program under Grant No. DGE 2146752.
Any opinions, findings, and conclusions or recommendations
expressed in this material are those of the author(s) and do not
necessarily reflect the views of the National Science Foundation.

\section*{Author Declarations}

\subsection*{Conflict of Interest}
The authors have no conflicts to disclose. 

\subsection*{Data Availability}
The data that support the findings of this study are available from the corresponding author upon reasonable request.

\bibliographystyle{achemso}
\bibliography{main}

\end{document}

\centering{\textbf{\Large{
Supplementary Material: Studying excited-state-specific perturbation theory
on the Thiel set}
}}
% Force line breaks with \\

\centering{Rachel Clune,\textsuperscript{a,1} Jacqueline A. R. Shea,\textsuperscript{b,}\footnote{These authors contributed equally to this work.} Tarini S. Hardikar,\textsuperscript{a} Harrison Tuckman,\textsuperscript{b} Eric Neuscamman\textsuperscript{a,b,}\footnote{eneuscamman@berkeley.edu}}
\\
\textsuperscript{a}Department of Chemistry, University of California, Berkeley, California 94720, USA\\
\textsuperscript{b}VeriSIM Life, San Francisco, California 94104, USA\\
%\email{eneuscamman@berkeley.edu}
\textsuperscript{c}Chemical Sciences Division, Lawrence Berkeley National Laboratory, Berkeley, CA, 94720, USA

\clearpage
% TODO: add level shfit figures and tables

%\section*{Acetamide}
%\input{ESMP2_vs_TBE/acetamide_vs_TBE_errors}
%\input{overlaps/acetamide_overlaps}
%\input{CI_amps/acetamide_CI_amps_table}
%\input{SA-CAS percentages/acetamide_SA_CAS_percentages}
%\input{ESMP2_amps/acetamide_vs_TBE_errors_w_amps}

%\clearpage
%\section*{Formaldehyde}
%\input{CI_amps/formaldehyde_CI_amps_table}

% This is without any states removed
\begin{table*}[htbp]
\caption{Mean unsigned errors and standard deviations for
singlet excitation energies in eV.
All states are included, except in the case
of ESMP2 and
$\epsilon$-ESMP2, which by necessity exclude
states that lack ESMF solutions
(gray rows in Table I).
\label{tab:wfn_all}
}
\resizebox{\textwidth}{!}{
\begin{tabular}{|l|c c c c c c c|}
\hline
\hspace{2mm} & 
\hspace{2mm} SA-CASPT2 & 
\hspace{2mm} MS-CASPT2 & 
\hspace{2mm} CC2 & 
\hspace{2mm} EOM-CCSD & 
\hspace{2mm} CC3 & 
\hspace{2mm} ESMP2 & 
\hspace{2mm} $\epsilon$-ESMP2 \\ \hline
\hspace{0.0mm} Ketones and amides & 0.20 $\pm$ 0.18 & 0.02 $\pm$ 0.05 & 0.29 $\pm$ 0.26 & 0.45 $\pm$ 0.38 & 0.26 $\pm$ 0.31 & 0.39 $\pm$ 0.37 & 0.17 $\pm$ 0.16 \\ 
\hspace{0.0mm} Conjugated polyenes & 0.15 $\pm$ 0.09 & 0.31 $\pm$ 0.25 & 0.76 $\pm$ 0.59 & 0.88 $\pm$ 0.46 & 0.44 $\pm$ 0.15 & 1.33 $\pm$ 1.36 & 0.31 $\pm$ 0.26 \\ 
\hspace{0.0mm} Conjugated rings & 0.33 $\pm$ 0.20 & 0.02 $\pm$ 0.05 & 0.29 $\pm$ 0.18 & 0.45 $\pm$ 0.25 & 0.18 $\pm$ 0.12 & 0.67 $\pm$ 0.43 & 0.21 $\pm$ 0.22 \\ 
\hspace{0.0mm} Heterocycles & 0.34 $\pm$ 0.21 & 0.10 $\pm$ 0.13 & 0.28 $\pm$ 0.19 & 0.42 $\pm$ 0.19 & 0.23 $\pm$ 0.16 & 0.71 $\pm$ 0.43 & 0.17 $\pm$ 0.14 \\ 
\hspace{0.0mm} Nucleobases & 0.40 $\pm$ 0.35 & 0.14 $\pm$ 0.10 & 0.20 $\pm$ 0.12 & 0.46 $\pm$ 0.31 & 0.15 $\pm$ 0.09 & 0.68 $\pm$ 0.52 & 0.19 $\pm$ 0.15 \\ \hline
\hspace{0.0mm} All & 0.31 $\pm$ 0.24 & 0.10 $\pm$ 0.13 & 0.30 $\pm$ 0.26 & 0.47 $\pm$ 0.30 & 0.23 $\pm$ 0.19 & 0.68 $\pm$ 0.58 & 0.19 $\pm$ 0.17 \\ \hline
\end{tabular}}
\end{table*}

\begin{table*}[htbp]
\caption{Mean unsigned errors and standard deviations for
singlet excitation energies in eV.
States without ESMF solutions
(gray rows in Table I)
are excluded.
%Mean unsigned errors and standard deviations in eV for molecular structure groups. All states that ESMF was unable to converge to were excluded, see the gray-highlighted states in Table 1.
\label{tab:wfn_gray}
}
\resizebox{\textwidth}{!}{
\begin{tabular}{|l|c c c c c c c|}
\hline
\hspace{2mm} & 
\hspace{2mm} SA-CASPT2 & 
\hspace{2mm} MS-CASPT2 & 
\hspace{2mm} CC2 & 
\hspace{2mm} EOM-CCSD & 
\hspace{2mm} CC3 & 
\hspace{2mm} ESMP2 & 
\hspace{2mm} $\epsilon$-ESMP2 \\ \hline
\hspace{0.0mm} Ketones and amides & 0.20 $\pm$ 0.18 & 0.02 $\pm$ 0.05 & 0.29 $\pm$ 0.26 & 0.45 $\pm$ 0.38 & 0.26 $\pm$ 0.31 & 0.39 $\pm$ 0.37 & 0.17 $\pm$ 0.16 \\ 
\hspace{0.0mm} Conjugated polyenes & 0.15 $\pm$ 0.09 & 0.31 $\pm$ 0.25 & 0.76 $\pm$ 0.59 & 0.88 $\pm$ 0.46 & 0.44 $\pm$ 0.15 & 1.33 $\pm$ 1.36 & 0.31 $\pm$ 0.26 \\ 
\hspace{0.0mm} Conjugated rings & 0.31 $\pm$ 0.19 & 0.02 $\pm$ 0.05 & 0.27 $\pm$ 0.17 & 0.41 $\pm$ 0.20 & 0.18 $\pm$ 0.12 & 0.67 $\pm$ 0.43 & 0.21 $\pm$ 0.22 \\ 
\hspace{0.0mm} Heterocycles & 0.34 $\pm$ 0.21 & 0.10 $\pm$ 0.13 & 0.28 $\pm$ 0.19 & 0.42 $\pm$ 0.19 & 0.23 $\pm$ 0.16 & 0.71 $\pm$ 0.43 & 0.17 $\pm$ 0.14 \\ 
\hspace{0.0mm} Nucleobases & 0.41 $\pm$ 0.37 & 0.15 $\pm$ 0.10 & 0.21 $\pm$ 0.13 & 0.47 $\pm$ 0.33 & 0.17 $\pm$ 0.08 & 0.68 $\pm$ 0.52 & 0.19 $\pm$ 0.15 \\ \hline
\hspace{0.0mm} All & 0.31 $\pm$ 0.24 & 0.10 $\pm$ 0.14 & 0.30 $\pm$ 0.27 & 0.47 $\pm$ 0.30 & 0.23 $\pm$ 0.19 & 0.68 $\pm$ 0.58 & 0.19 $\pm$ 0.17 \\ \hline
\end{tabular}}
\end{table*}

\clearpage
\begin{table*}[htbp]
\caption{Mean unsigned errors and standard deviations for
singlet excitation energies in eV.
All states are included, except in the case
of ESMP2 and
$\epsilon$-ESMP2, which by necessity exclude
states that lack ESMF solutions
(gray rows in Table I).
\label{tab:td_dft_all}
}
\resizebox{\textwidth}{!}{
\begin{tabular}{|l|c c c c c c|}
\hline
\hspace{6mm} & 
\hspace{6mm} BP86 & 
\hspace{6mm} B3LYP & 
\hspace{6mm} BHLYP & 
\hspace{6mm} DFT/MRCI & 
\hspace{6mm} ESMP2 & 
\hspace{6mm} $\epsilon$-ESMP2 \\ 
\hline
\hspace{0.0mm} Ketones and amides & 0.55 $\pm$ 0.35 & 0.29 $\pm$ 0.19 & 0.35 $\pm$ 0.44 & 0.34 $\pm$ 0.21 & 0.39 $\pm$ 0.37 & 0.17 $\pm$ 0.16 \\ 
\hspace{0.0mm} Conjugated polyenes & 0.38 $\pm$ 0.30 & 0.40 $\pm$ 0.19 & 0.70 $\pm$ 0.62 & 0.27 $\pm$ 0.14 & 1.33 $\pm$ 1.36 & 0.31 $\pm$ 0.26 \\ 
\hspace{0.0mm} Conjugated rings & 0.47 $\pm$ 0.33 & 0.35 $\pm$ 0.19 & 0.42 $\pm$ 0.36 & 0.23 $\pm$ 0.23 & 0.67 $\pm$ 0.43 & 0.21 $\pm$ 0.22 \\ 
\hspace{0.0mm} Heterocycles & 0.43 $\pm$ 0.29 & 0.20 $\pm$ 0.18 & 0.50 $\pm$ 0.26 & 0.18 $\pm$ 0.12 & 0.71 $\pm$ 0.43 & 0.17 $\pm$ 0.14 \\ 
\hspace{0.0mm} Nucleobases & 0.82 $\pm$ 0.30 & 0.47 $\pm$ 1.14 & 0.57 $\pm$ 0.27 & 0.16 $\pm$ 0.13 & 0.68 $\pm$  0.52 & 0.19 $\pm$ 0.15 \\ \hline
\hspace{0.0mm} All & 0.53 $\pm$ 0.34 & 0.31 $\pm$ 0.52 & 0.49 $\pm$ 0.35 & 0.22 $\pm$ 0.17 & 0.68 $\pm$ 0.58 & 0.19 $\pm$ 0.17 \\ \hline
\end{tabular}}
\end{table*}

\begin{table*}[htbp]
\caption{Mean unsigned errors and standard deviations for
singlet excitation energies in eV.
States without ESMF solutions
(gray rows in Table I)
are excluded.
\label{tab:td_dft_gray}
}
\resizebox{\textwidth}{!}{
\begin{tabular}{|l|c c c c c c|}
\hline
\hspace{6mm} & 
\hspace{6mm} BP86 & 
\hspace{6mm} B3LYP & 
\hspace{6mm} BHLYP & 
\hspace{6mm} DFT/MRCI & 
\hspace{6mm} ESMP2 & 
\hspace{6mm} $\epsilon$-ESMP2 \\ 
\hline
\hspace{0.0mm} Ketones and amides & 0.55 $\pm$ 0.35 & 0.29 $\pm$ 0.19 & 0.35 $\pm$ 0.44 & 0.34 $\pm$ 0.21 & 0.39 $\pm$ 0.37 & 0.17 $\pm$ 0.16 \\ 
\hspace{0.0mm} Conjugated polyenes & 0.38 $\pm$ 0.30 & 0.40 $\pm$ 0.19 & 0.70 $\pm$ 0.62 & 0.27 $\pm$ 0.14 & 1.33 $\pm$ 1.36 & 0.31 $\pm$ 0.26 \\ 
\hspace{0.0mm} Conjugated rings & 0.47 $\pm$ 0.34 & 0.36 $\pm$ 0.19 & 0.39 $\pm$ 0.34 & 0.20 $\pm$ 0.21 & 0.67 $\pm$ 0.43 & 0.21 $\pm$ 0.22 \\ 
\hspace{0.0mm} Heterocycles & 0.43 $\pm$ 0.29 & 0.20 $\pm$ 0.18 & 0.50 $\pm$ 0.26 & 0.18 $\pm$ 0.12 & 0.71 $\pm$ 0.43 & 0.17 $\pm$ 0.14 \\ 
\hspace{0.0mm} Nucleobases & 0.83 $\pm$ 0.30 & 0.50 $\pm$ 1.20 & 0.57 $\pm$ 0.29 & 0.15 $\pm$ 0.12 & 0.68 $\pm$  0.52 & 0.19 $\pm$ 0.15 \\ \hline
\hspace{0.0mm} All & 0.52 $\pm$ 0.34 & 0.31 $\pm$ 0.52 & 0.48 $\pm$ 0.35 & 0.21 $\pm$ 0.17 & 0.68 $\pm$ 0.58 & 0.19 $\pm$ 0.17 \\ \hline
\end{tabular}}
\end{table*}

\clearpage
\begin{table*}[htbp]
\caption{Mean unsigned errors and standard deviations for
singlet excitation energies in eV.
All states are included, except in the case
of ESMP2 and
$\epsilon$-ESMP2, which by necessity exclude
states that lack ESMF solutions
(gray rows in Table I).
%Mean unsigned errors in eV for molecular groupings based on pi system size. All states have been included in the calculation of these averages.
\label{tab:wfnc_pi_all}
}
\resizebox{\textwidth}{!}{
\begin{tabular}{|l|c c c c c c c|}
\hline
\multicolumn{1}{|l|}{$\pi$ system size} & 
\hspace{2mm} SA-CASPT2 & 
\hspace{2mm} MS-CASPT2 & 
\hspace{2mm} CC2 & 
\hspace{2mm} EOM-CCSD & 
\hspace{2mm} CC3 & 
\hspace{2mm} ESMP2 & 
\hspace{2mm} $\epsilon$-ESMP2 \\ \hline
\hspace{0.0mm} 2 & 0.19 $\pm$ 0.17 & 0.11 $\pm$ 0.24 & 0.32 $\pm$ 0.27 & 0.39 $\pm$ 0.36 & 0.28 $\pm$ 0.34 & 0.21 $\pm$ 0.14 & 0.17 $\pm$ 0.16 \\ 
\hspace{0.0mm} 3 & 0.12 $\pm$ 0.11 & 0.00 $\pm$ 0.00 & 0.28 $\pm$ 0.26 & 0.40 $\pm$ 0.43 & 0.30 $\pm$ 0.32 & 0.16 $\pm$ 0.18 & 0.13 $\pm$ 0.09 \\ 
\hspace{0.0mm} 4 & 0.25 $\pm$ 0.15 & 0.15 $\pm$ 0.15 & 0.57 $\pm$ 0.42 & 0.62 $\pm$ 0.24 & 0.28 $\pm$ 0.10 & 1.05 $\pm$ 0.99 & 0.23 $\pm$ 0.18 \\ 
\hspace{0.0mm} 5 & 0.41 $\pm$ 0.12 & 0.09 $\pm$ 0.10 & 0.37 $\pm$ 0.18 & 0.41 $\pm$ 0.17 & 0.19 $\pm$ 0.15 & 0.48 $\pm$ 0.50 & 0.16 $\pm$ 0.14 \\ 
\hspace{0.0mm} 6 & 0.31 $\pm$ 0.22 & 0.12 $\pm$ 0.13 & 0.30 $\pm$ 0.28 & 0.46 $\pm$ 0.27 & 0.24 $\pm$ 0.18 & 0.80 $\pm$ 0.50 & 0.20 $\pm$ 0.21 \\ 
\hspace{0.0mm} 8 & 0.33 $\pm$ 0.25 & 0.10 $\pm$ 0.11 & 0.27 $\pm$ 0.28 & 0.56 $\pm$ 0.36 & 0.19 $\pm$ 0.12 & 0.86 $\pm$ 0.72 & 0.22 $\pm$ 0.17 \\ 
\hspace{0.0mm} 10 & 0.49 $\pm$ 0.33 & 0.03 $\pm$ 0.07 & 0.24 $\pm$ 0.15 & 0.44 $\pm$ 0.23 & 0.17 $\pm$ 0.09 & 0.69 $\pm$ 0.41 & 0.16 $\pm$ 0.10 \\ \hline
\multicolumn{1}{|l|}{5 or less} & 0.24 $\pm$ 0.18 & 0.09 $\pm$ 0.17 & 0.36 $\pm$ 0.27 & 0.43 $\pm$ 0.31 & 0.26 $\pm$ 0.26 & 0.40 $\pm$ 0.52 & 0.17 $\pm$ 0.14 \\ 
\multicolumn{1}{|l|}{6 or more} & 0.34 $\pm$ 0.25 & 0.10 $\pm$ 0.10 & 0.28 $\pm$ 0.26 & 0.49 $\pm$ 0.30 & 0.23 $\pm$ 0.17 & 0.80 $\pm$ 0.56 & 0.20 $\pm$ 0.18 \\ \hline
\end{tabular}}
\end{table*}

\begin{table*}[htbp]
\caption{Mean unsigned errors and standard deviations for
singlet excitation energies in eV.
States without ESMF solutions
(gray rows in Table I)
are excluded.
%Mean unsigned errors in eV for molecular groupings based on pi system size. The states that ESMF did not converge to have been omitted from these values, see the gray-highlighted states in Table 1.
\label{tab:wfnc_pi_gray}
}
\resizebox{\textwidth}{!}{
\begin{tabular}{|l|c c c c c c c|}
\hline
\multicolumn{1}{|l|}{$\pi$ system size} & 
\hspace{2mm} SA-CASPT2 & 
\hspace{2mm} MS-CASPT2 & 
\hspace{2mm} CC2 & 
\hspace{2mm} EOM-CCSD & 
\hspace{2mm} CC3 & 
\hspace{2mm} ESMP2 & 
\hspace{2mm} $\epsilon$-ESMP2 \\ \hline
\hspace{0.0mm} 2 & 0.19 $\pm$ 0.17 & 0.11 $\pm$ 0.24 & 0.32 $\pm$ 0.27 & 0.39 $\pm$ 0.36 & 0.28 $\pm$ 0.34 & 0.21 $\pm$ 0.14 & 0.17 $\pm$ 0.16 \\ 
\hspace{0.0mm} 3 & 0.12 $\pm$ 0.11 & 0.00 $\pm$ 0.00 & 0.28 $\pm$ 0.26 & 0.40 $\pm$ 0.43 & 0.30 $\pm$ 0.32 & 0.16 $\pm$ 0.18 & 0.13 $\pm$ 0.09 \\ 
\hspace{0.0mm} 4 & 0.15 $\pm$ 0.15 & 0.15 $\pm$ 0.15 & 0.57 $\pm$ 0.42 & 0.62 $\pm$ 0.24 & 0.28 $\pm$ 0.10 & 1.05 $\pm$ 0.99 & 0.23 $\pm$ 0.18 \\ 
\hspace{0.0mm} 5 & 0.41 $\pm$ 0.12 & 0.09 $\pm$ 0.10 & 0.37 $\pm$ 0.18 & 0.41 $\pm$ 0.17 & 0.19 $\pm$ 0.15 & 0.48 $\pm$ 0.50 & 0.16 $\pm$ 0.14 \\ 
\hspace{0.0mm} 6 & 0.31 $\pm$ 0.22 & 0.12 $\pm$ 0.13 & 0.30 $\pm$ 0.28 & 0.46 $\pm$ 0.27 & 0.24 $\pm$ 0.18 & 0.80 $\pm$ 0.50 & 0.20 $\pm$ 0.21 \\ 
\hspace{0.0mm} 8 & 0.32 $\pm$ 0.25 & 0.10 $\pm$ 0.11 & 0.27 $\pm$ 0.29 & 0.56 $\pm$ 0.29 & 0.20 $\pm$ 0.12 & 0.86 $\pm$ 0.72 & 0.22 $\pm$ 0.17 \\ 
\hspace{0.0mm} 10 & 0.48 $\pm$ 0.34 & 0.04 $\pm$ 0.07 & 0.20 $\pm$ 0.08 & 0.39 $\pm$ 0.13 & 0.17 $\pm$ 0.10 & 0.69 $\pm$ 0.41 & 0.16 $\pm$ 0.10 \\ \hline
\multicolumn{1}{|l|}{5 or less} & 0.24 $\pm$ 0.18 & 0.09 $\pm$ 0.17 & 0.36 $\pm$ 0.27 & 0.43 $\pm$ 0.31 & 0.26 $\pm$ 0.26 & 0.40 $\pm$ 0.52 & 0.17 $\pm$ 0.14 \\ 
\multicolumn{1}{|l|}{6 or more} & 0.34 $\pm$ 0.25 & 0.10 $\pm$ 0.10 & 0.27 $\pm$ 0.26 & 0.48 $\pm$ 0.29 & 0.22 $\pm$ 0.15 & 0.80 $\pm$ 0.56 & 0.20 $\pm$ 0.18 \\ \hline
\end{tabular}}
\end{table*}

%\begin{figure}
%    \caption{Singles percentages calculated from the first order ESMP wave function versus the corresponding %SA-CASSCF singles percentages. R\textsuperscript{2} value of 0.282.}
%    \includegraphics[scale=0.7]{plots/singles_norm_vs_SA_CASSCF.pdf}
%    \label{singles_CASSCF}
%\end{figure}
%
%\begin{figure}
%    \caption{Singles percentages calculated from the first order lvl-ESMP wave function versus the %corresponding lvl-ESMP2 errors in eV. R\textsuperscript{2} value of 0.004.}
%    \includegraphics[scale=0.7]{plots/singles_norm_vs_lvl_ESMP2_error.pdf}
%    \label{singles_lvl}
%\end{figure}

\clearpage

\section*{Level shift examples}

Below are plots showing the effects of different level shifts on the excitation energies of representative molecules from each of the groups in the benchmark.  Note that the vertical axis scale differs from figure to figure.

\begin{figure}[ht]
    \centering
    \includegraphics[scale=0.7]{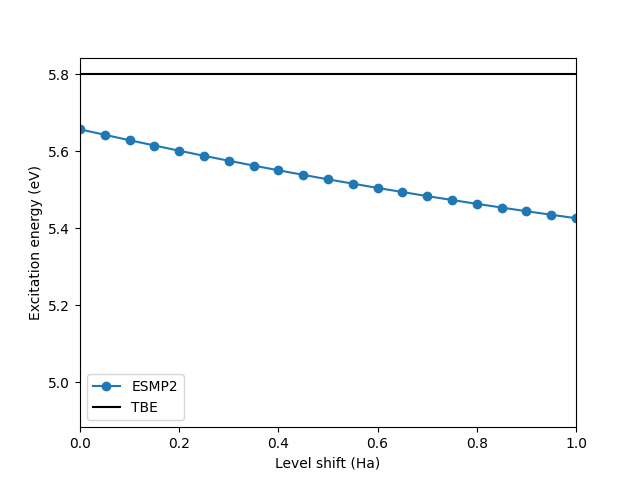}
    \caption{Predicted excitation energies for the 1$^1$A$^{\prime\prime}$ state of acetamide for different values of the level shift in $\varepsilon$-ESMP2. Note that the change in the excitation energy of this state between a level shift of 1.0 Ha and 0.0 Ha is about 0.1 eV, showing that the level shift had very little impact on the accuracy of the ESMP2 prediction.}
    \label{fig:acetone_11Ad_SI}
\end{figure}
\begin{figure}[ht]
    \centering
    \includegraphics[scale=0.7]{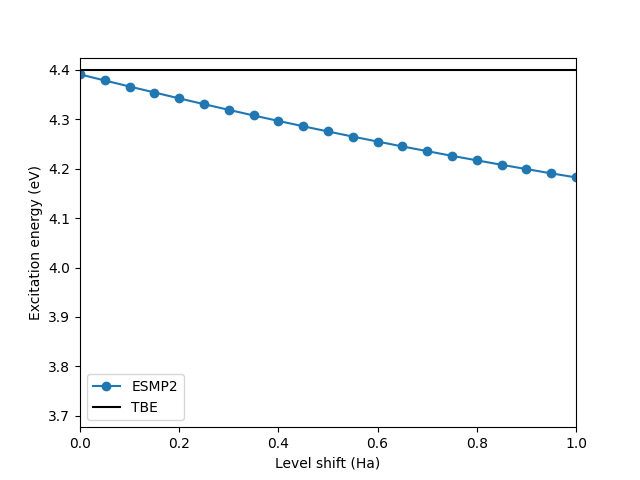}
    \caption{Predicted excitation energies for the 2$^1$A$^{\prime}$ state of acetamide for different values of the level shift in $\varepsilon$-ESMP2. The change in the predicted excitation energy across the range of shifts is only about 0.2 eV, showing that the level shift has very little impact on the accuracy of the prediction. }
    \label{fig:acetamide_21Ap_SI}
\end{figure}

\begin{figure}[ht]
    \centering
    \includegraphics[scale=0.7]{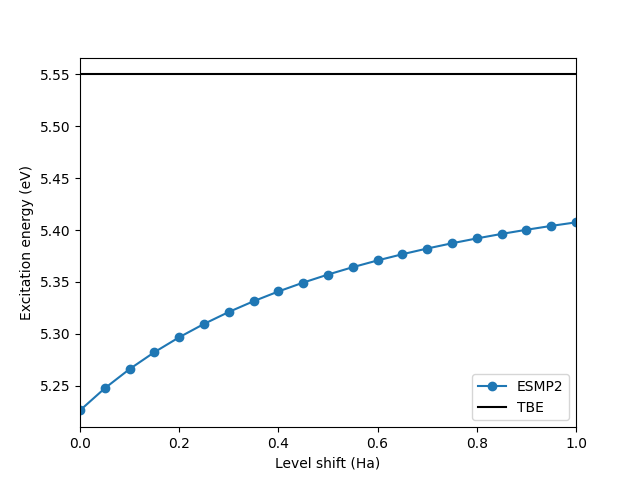}
    \caption{Predicted excitation energies for the 1$^1$B$_2$ state of cyclopentadiene for different values of the level shift in $\varepsilon$-ESMP2. The change of predicted excitation energy for the shifts shown here is only 0.2 eV, showing that the predicted energy is not very sensitive to the choice of shift. }
    \label{fig:cyclopentadiene_11B2_SI}
\end{figure}

\begin{figure}[ht]
    \centering
    \includegraphics[scale=0.7]{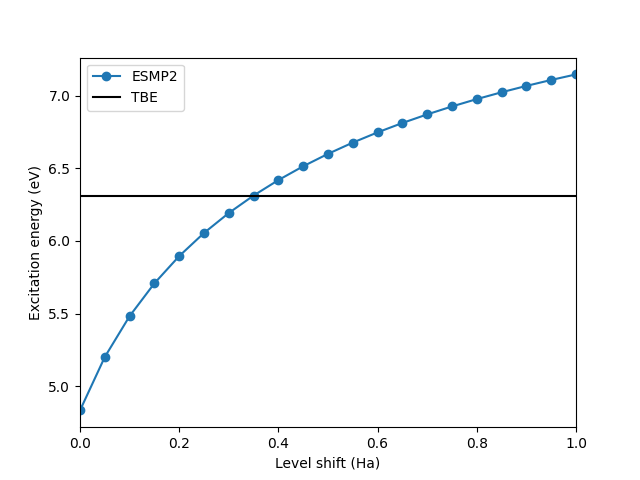}
    \caption{Predicted excitation energies for the 2$^1$A$_1$ state of cyclopentadiene for different values of the level shift in $\varepsilon$-ESMP2. This state, likely because it has some doubly excited character, was very impacted by the addition of the level shift to the ESMP2 method.}
    \laåbel{fig:cyclopentadiene_21A1_SI}
\end{figure}

\begin{figure}[ht]
    \centering
    \includegraphics[scale=0.7]{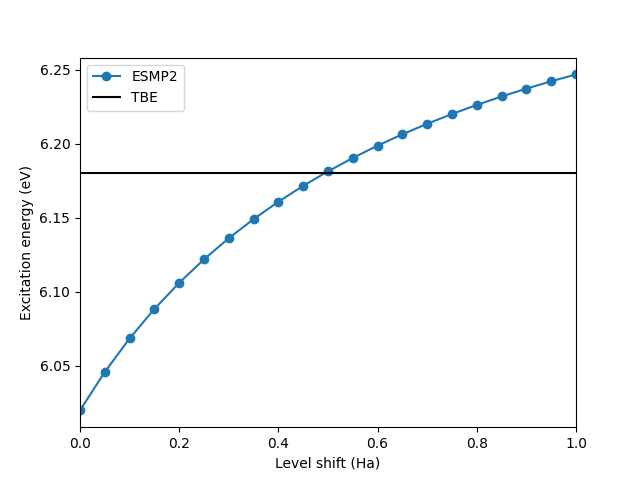}
    \caption{Predicted excitation energies for the 1$^1$B$_u$ state of butadiene for different values of the level shift in $\varepsilon$-ESMP2. The excitation energy only changes by 0.2eV across the range of tested level shifts, showing that this state is not very sensitive to the choice of shift. }
    \label{fig:butadiene_11Bu_SI}
\end{figure}

\begin{figure}[ht]
    \centering
    \includegraphics[scale=0.7]{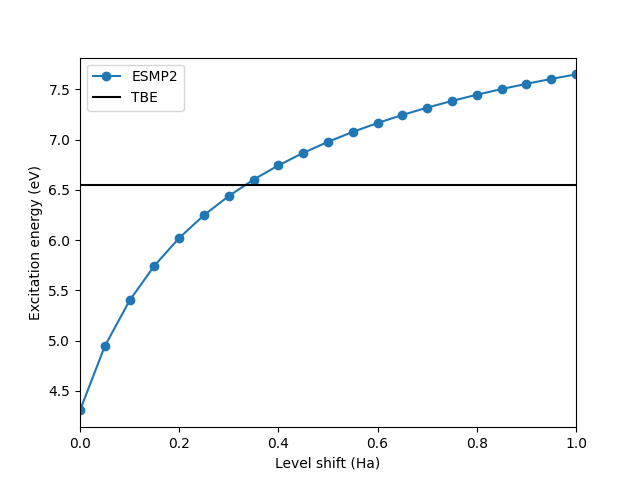}
    \caption{Predicted excitation energies for the 2$^1$A$_g$ state of butadiene for different values of the level shift in $\varepsilon$-ESMP2. Likely due to its partly doubly excited character, this state was highly sensitive to the level shift value.}
    \label{fig:butadiene_21Ag_SI}
\end{figure}

\begin{figure}[ht]
    \centering
    \includegraphics[scale=0.7]{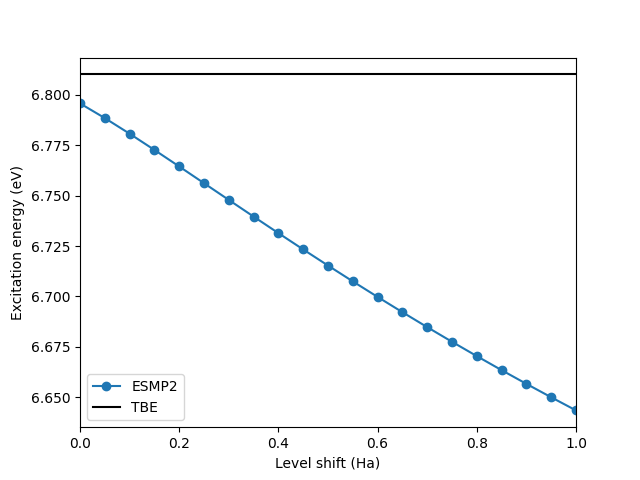}
    \caption{Predicted excitation energies for the 1$^1$A$^{\prime\prime}$ state of imidazole for different values of the level shift in $\varepsilon$-ESMP2. This state was only slightly impacted by the addition of the level shift, as the range of predicted excitation energies only varies by 0.15 eV. }
    \label{fig:imidazole_11Ad_SI}
\end{figure}

\begin{figure}[ht]
    \centering
    \includegraphics[scale=0.7]{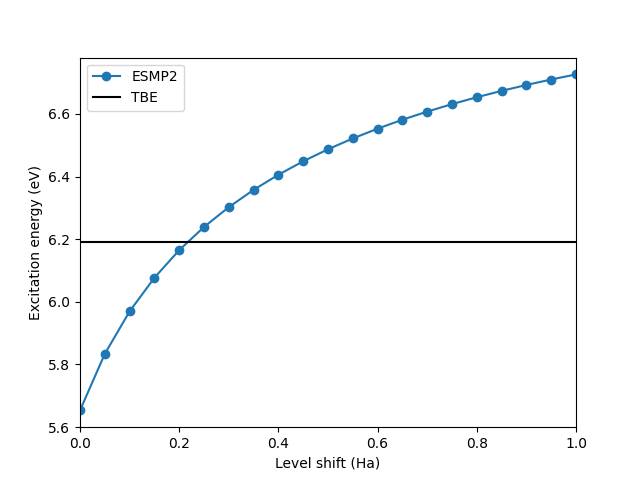}
    \caption{Predicted excitation energies for the 2$^1$A$^{\prime}$ state of imidazole for different values of the level shift in $\varepsilon$-ESMP2. This state is a good example of a state in a $\pi$ system that is somewhat sensitive to regularization.}
    \label{fig:imidazole_21Ap_SI}
\end{figure}

\begin{figure}[ht]
    \centering
    \includegraphics[scale=0.7]{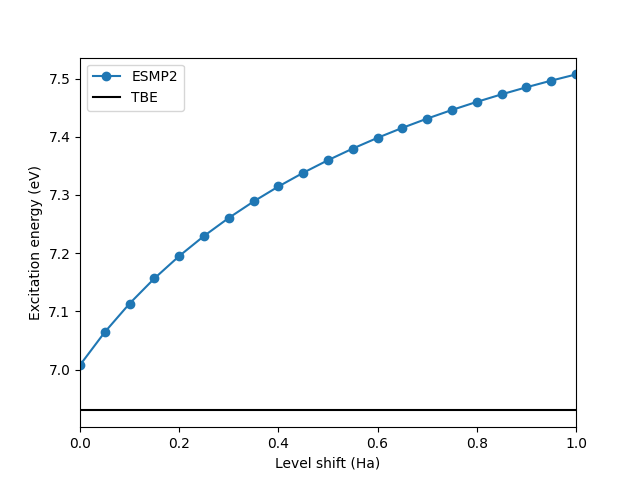}
    \caption{Predicted excitation energies for the 3$^1$A$^{\prime}$ state of imidazole for different values of the level shift in $\varepsilon$-ESMP2. The addition of the level shift to ESMP2 had a large impact on the predicted excitation energy for this state, and it is a good example of a case where regularization does not improve accuracy.}
    \label{fig:imidazole_31Ap_SI}
\end{figure}

\begin{figure}[ht]
    \centering
    \includegraphics[scale=0.7]{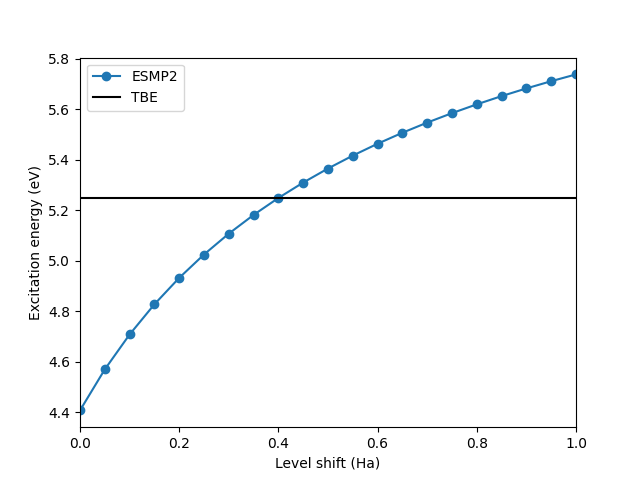}
    \caption{Predicted excitation energies for the 2$^1$A$^{\prime}$ state of adenine for different values of the level shift in $\varepsilon$-ESMP2. This is another example of a $\pi$ system excitation that is sensitive to regularization.}
    \label{fig:adenine_21Ap_SI}
\end{figure}

\begin{figure}[ht]
    \centering
    \includegraphics[scale=0.7]{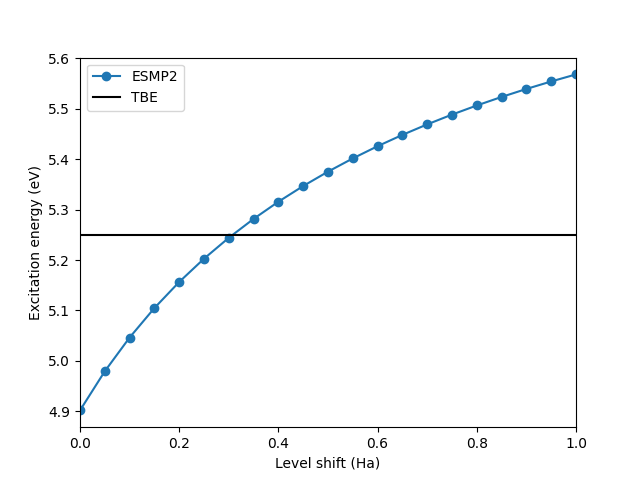}
    \caption{Predicted excitation energies for the 3$^1$A$^{\prime}$ state of adenine for different values of the level shift in $\varepsilon$-ESMP2. Similar to the 2$^1$A$^{\prime}$ state above, adding regularization makes a significant difference here.}
    \label{fig:adenine_31Ap_SI}
\end{figure}

\begin{figure}[ht]
    \centering
    \includegraphics[scale=0.7]{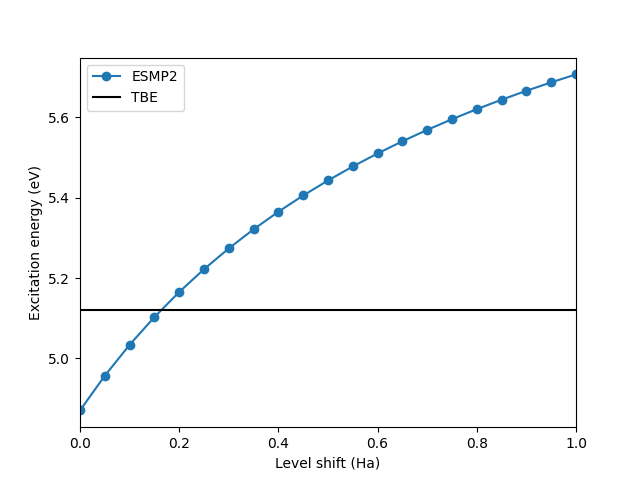}
    \caption{Predicted excitation energies for the 1$^1$A$^{\prime\prime}$ state of adenine for different values of the level shift in $\varepsilon$-ESMP2. }
    \label{fig:adenine_11Ad_SI}
\end{figure}

\begin{figure}[ht]
    \centering
    \includegraphics[scale=0.7]{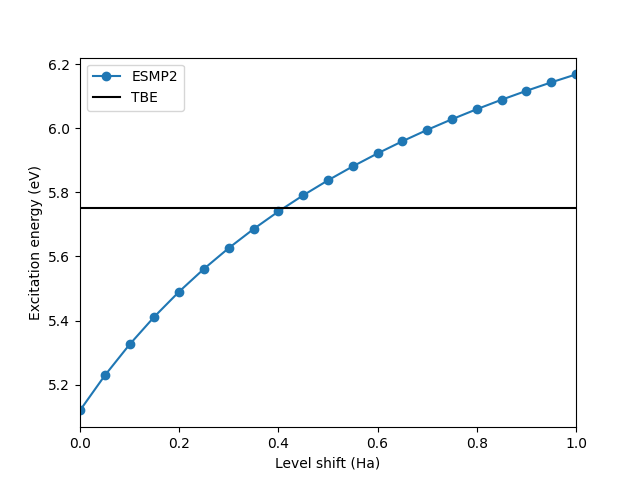}
    \caption{Predicted excitation energies for the 2$^1$A$^{\prime\prime}$ state of adenine for different values of the level shift in $\varepsilon$-ESMP2. Like the other adenine states studied here, the addition of a level shift makes a significant difference. }
    \label{fig:adenine_21Ad_SI}
\end{figure}

\clearpage

%\input{main paper tables/all_molecules_all_methods}
% Trying to do the above with for loops, this should be interesting
\foreach \n in {Formaldehyde,Acetone,Benzoquinone,Formamide,Acetamide,Propanamide,Ethene,Butadiene,Hexatriene,Octatetraene,Cyclopropene,Cyclopentadiene,Norbornadiene,Benzene,Naphthalene,Furan,Pyrrole,Imidazole,Pyridine,Pyrazine,Pyrimidine,Pyridazine,Triazine,Tetrazine,Cytosine,Thymine,Uracil,Adenine}{
\section*{\n}
\input{eV values/\n_all_methods}
\input{ESMP2_vs_TBE/\n_vs_TBE_errors}
\input{ESMP2_amps/\n_vs_TBE_errors_w_amps}
\input{SA-CAS percentages/\n_SA_CAS_percentages}
\input{CI_amps/\n_CI_amps_table}
\clearpage
}